\documentclass[10pt,aps,prl,twocolumn,groupedaddress,superscriptaddress,showkeys]{revtex4-2}
\usepackage{amsmath}
\usepackage{graphicx}
\DeclareGraphicsExtensions{.pdf,.eps,.png,.jpg,.mps}
\DeclareUnicodeCharacter{2061}{}
\usepackage{epstopdf}
\usepackage{color}
\usepackage{float}
\usepackage{soul}
\usepackage{amssymb}
\usepackage[normalem]{ulem}
\usepackage{todonotes}
\usepackage{booktabs}
\usepackage{adjustbox}
\usepackage{hyperref}

\begin{document}
\title{Graphene Electro-Absorption Modulators for Energy-Efficient and High-Speed Optical Transceivers}
\author{M. Tiberi}
\affiliation{Cambridge Graphene Centre, University of Cambridge, 9 JJ Thomson Avenue, CB3 0FA, Cambridge, UK}
\author{A. Montanaro}
\affiliation{Photonic Networks and Technologies Lab, CNIT, 56124, Pisa, Italy}
\author{C. Wen}
\author{J. Zhang}
\author{O. Balci}
\author{S. M. Shinde}
\author{S. Sharma}
\author{A. Meersha}
\author{H. Shekhar}
\author{J. E. Muench}
\affiliation{Cambridge Graphene Centre, University of Cambridge, 9 JJ Thomson Avenue, CB3 0FA, Cambridge, UK}
\author{B. R. Conran}
\author{K. B. K. Teo}
\affiliation{AIXTRON Ltd, Buckingway Business Park, Anderson Rd, Swavesey, CB24 4FQ, Cambridge, UK}
\author{M. Ebert}
\author{X. Yan}
\author{Y. Tran}
\author{M. Banakar}
\author{C. Littlejohns}
\author{G. T. Reed}
\affiliation{Optoelectronics Research Centre, University of Southampton, SO17 1BJ, Southampton, UK}
\author{M. Romagnoli}
\affiliation{Photonic Networks and Technologies Lab, CNIT, 56124, Pisa, Italy}
\affiliation{CamGraPhIC srl, Pisa, Italy}
\author{A. Ruocco}
\affiliation{Cambridge Graphene Centre, University of Cambridge, 9 JJ Thomson Avenue, CB3 0FA, Cambridge, UK}
\author{V. Sorianello}
\affiliation{Photonic Networks and Technologies Lab, CNIT, 56124, Pisa, Italy}
\author{A. C. Ferrari}
\email{acf26@eng.cam.ac.uk}
\affiliation{Cambridge Graphene Centre, University of Cambridge, 9 JJ Thomson Avenue, CB3 0FA, Cambridge, UK}
\thispagestyle{empty}
\pagestyle{empty}
\begin{abstract}
The increasing demand for energy-efficient hardware for artificial intelligence (AI) and data centres requires integrated photonic solutions delivering optical transceivers with Tbit/s data rates and energy consumption$<$1pJ/bit. Here, we report double single-layer graphene electro-absorption modulators on Si optimized for energy-efficient and ultra-fast operation, demonstrating 67GHz bandwidth and 80Gbit/s data rate, in both O and C bands, using a fabrication tailored for wafer-scale integration. We measure a data rate$\sim$1.6 times larger than previously reported for graphene. We scale the modulator's active area down to 22$\mu$m$^2$, achieving a dynamic power consumption$\sim$58fJ/bit, $\sim$3 times lower than previous graphene modulators and Mach-Zehnder modulators based on Si or lithium niobate. We show devices with$\sim$0.037dB/V$\mu$m modulation efficiency,$\sim$16 times better than previous demonstrations based on graphene. This paves the way to wafer-scale production of graphene modulators on Si useful for Tbit/s optical transceivers and energy-efficient AI.
\end{abstract}
\maketitle
Artificial intelligence (AI) and machine learning (ML) are generating an unprecedented demand for computational capacity\cite{Ossieur2025, EpochAI} and energy\cite{EthernetAlliance2024, IEAElectricity2024}. The underlying hardware is a network of processing units working in parallel\cite{Ossieur2025, Ishii2022}, which can be linked via optical interconnects and switches\cite{Maniotis2024,Cheng2018}. These networks rely on short-reach optical transceivers\cite{Ossieur2023} ranging from few cm to 2km\cite{Ossieur2025, Zhou2020}, with high capacity (Tbit/s\cite{Ossieur2025, EthernetAlliance2024, Ossieur2023}) to match processing units speed and reduce latency\cite{Ossieur2025}, i.e. the time a packet of data takes to travel between two points across a network connection\cite{IBMLatency}. Low energy consumption ($<$1pJ/bit\cite{Ossieur2025}) is also needed, as many ($>$1,000\cite{MukherjeeMedium2024}) processing units are used to train AI models\cite{Ossieur2025}.

In order to reach Tbit/s, one option is to develop devices and technologies that increase the baud rate of each optical lane\cite{Ossieur2025, EthernetAlliance2024}, i.e. the number of symbols transmitted per second in a communication channel\cite{Proakis2008}. In intensity-modulated direct-detection (IMDD) systems\cite{Proakis2008}, non-return-to-zero (NRZ)\cite{Proakis2008} and 4-level pulsed-amplitude-modulation (PAM-4)\cite{Proakis2008} schemes encode bits in the intensity of a laser\cite{Proakis2008}. In NRZ encoding, each symbol represents one bit\cite{Proakis2008}, so the baud rate equals the bit rate\cite{Proakis2008}. PAM-4 encoding uses 4 signal levels to represent 2 bits per symbol\cite{Proakis2008}, doubling the data rate, without increasing the baud rate\cite{Proakis2008}. Tbit/s data rates can be realised by combining multiple lanes in wavelength division multiplexing (WDM)\cite{Brackett1990}, with at least single-lane 50GBaud NRZ and PAM-4 data rates\cite{EthernetAlliance2024}. Coherent systems, alternative to IMDD, employ amplitude\cite{Kumar2014, Kikuchi2016}, phase\cite{Kumar2014, Kikuchi2016}, and polarization\cite{Kumar2014, Kikuchi2016} degrees of freedom to encode data\cite{Kumar2014,  Kikuchi2016}. Coherent systems increase spectral efficiency\cite{Kumar2014, Kikuchi2016}, i.e. the amount of transmitted data over a given wavelength\cite{Benedetto1999}. For this reason, coherent systems could be implemented in short-reach networks($<$100km\cite{Chagnon2019, IEEEStandard2023}) and data centers to increase spectral efficiency and overall transmission capacity.

Increasing single-lane baud rate and using IMDD schemes are still preferred for Tbit/s short-reach networks\cite{Zhou2020, Estaran2019}, particularly for intra-data-centre interconnections($<$2km\cite{Zhou2020}) and chip-to-chip interconnections, crucial for AI\cite{Ossieur2025}. This is because IMDD solutions require less energy than coherent ones, relying on digital signal processing (DSP) for phase\cite{Li2009} and polarization\cite{Li2009} management at the receiver. DSP increases both latency\cite{Ossieur2025} and energy consumption\cite{Ossieur2025}, being an additional data processing step. Ref.\cite{Cheng2019} showed that a coherent DSP chip consumes$\sim$25\% more than a IMDD one at the same capacity. DSP is now considered the major source of power consumption\cite{Nagarajan2021, Wei2015}, contributing$>$50\% to the total power budget in AI\cite{Nagarajan2021, Wei2015} and data centers optical interconnects\cite{Nagarajan2021, Wei2015}. Considering that data centers accounted$\sim$1-1.5\% of global electricity use in 2024\cite{EthernetAlliance2024}, projected to more than double by 2026\cite{IEAElectricity2024}, optical transceivers must not only increase the baud rate, but also reduce energy consumption from$\sim$100pJ/bit\cite{Lee2012} to$<$1pJ/bit\cite{Ossieur2025} in order to limit the environmental impact, hence reducing global emissions and achieving the worldwide goal of net zero by 2050\cite{IEAEnergy}. In 2020, global data centres were estimated to contribute$\sim$330M tonnes CO$_2$ annually\cite{IEADatacentresEmission}, while by 2030 this could surge to 2.5Bn tonnes\cite{ReutersEmissions}, corresponding to$\sim$7\% of the predicted total emissions.

Another source of latency and energy consumption in data centres is forward error correction (FEC)\cite{IntelFEC,CiscoFEC}, an error‐correction mechanism that adds redundant bits to the transmitted data to detect bit errors at the receiver\cite{IntelFEC,CiscoFEC} and reduce the bit error rate (BER)\cite{IntelFEC,CiscoFEC}, i.e. the number of bits altered per second due to noise\cite{Proakis2008}. Removing FEC would therefore reduce latency\cite{IntelFEC,CiscoFEC}. However, this is possible only if the inherent BER is already sufficiently low (BER$<$10$^{-12}$\cite{IntelFEC,CiscoFEC, IEEEBER}).

To achieve$<$1pJ/bit, the integrated photonic components, such as modulators, should have energy consumption$<$100fJ/bit\cite{Reed2010, Ho2013, Zhou2015}, and DSP use should be reduced\cite{Ossieur2025}. E.g., transceivers in linear pluggable optics with continuous-time linear equalizers (CTLE)\cite{Ossieur2025} or co-packaged optics\cite{Ossieur2025, Maniotis2024} employ DSP only in the electronic processing unit\cite{Ossieur2025}, but not in the optical transceiver\cite{Ossieur2025}. Ref.\cite{Li2024} reported a DSP-free Si transmitter using optical equalisation techniques, instead of DSP, achieving single-lane 308Gbit/s PAM-4\cite{Li2024} and$\sim$0.7pJ/bit\cite{Li2024}.

Transceivers based on NRZ modulation formats allow for$<$500m optical interconnects without DSP\cite{Verbist2018}. Nevertheless, fiber links$>$500m would require DSP, including in NRZ-based interconnects\cite{Cole2013,Forestieri2017}, primarily due to chromatic dispersion\cite{Wei2015, Forestieri2017, Plabst2022}, i.e. the wavelength-dependent variation of the optical signal phase over a propagation length\cite{Kumar2014}. For this reason, the 400 Gigabit Ethernet (GbE)\cite{IEEE200Gbps400GbpsETaskForce, IEEEStandardEthernet400GbpsAmendement} and 800 GbE\cite{IEEE400Gbps800GbpsETaskForce} standards are mainly based on PAM-4 modulation\cite{IEEEStandardEthernet400GbpsAmendement}, with one implementation adopting NRZ modulation formats (400GBASE-SR16\cite{IEEE200Gbps400GbpsETaskForce}). The adoption of DSP-free NRZ transceivers is an opportunity for increasing the baud rate to Tbit/s at low ($<$1pJ/bit) energy costs, but components with high baud rate ($>$50GBaud NRZ per lane\cite{EthernetAlliance2024}) and resilient to chromatic dispersion are needed to enable$>$500m interconnects without DSP.

Electro-absorption modulators (EAMs) modulate the intensity of an optical field by varying a material's optical absorption with an electric field\cite{Soref1987}. EAMs are crucial in IMDD systems\cite{Cheng2019} because they modulate light intensity, and could also be used in coherent systems, such as in-phase and quadrature (IQ) modulators\cite{Wei2018, Sorianello2023}. Important parameters for EAMs are the maximum static extinction ratio ER=10log(P$_{max}$/P$_{min}$), insertion loss IL=$\alpha$L, symbol rate (GBaud), and energy consumption per bit E$_{bit}$=$\frac{CV_{pp}^2}{4}$\cite{Miller2012}. P$_{max}$ and P$_{min}$ are the maximum and minimum transmitted optical powers, $\alpha$ is the absorption coefficient, $C$ is the capacitance, V$_{pp}$ is the peak-to-peak voltage applied to the modulator, and $L$ is the modulator's length. The dynamic ER, i.e. the ER during modulated transmission, depends on DC bias and V$_{pp}$ used to drive the modulator. Another key parameter is the electro-optic bandwidth (EO-BW), i.e. the cut-off frequency at which EO modulation is driven by an electric signal at a specific power loss, typically 3dB\cite{Reed2010}.

Combining the above parameters gives the following figures of merit (FOM): modulation efficiency normalized by length FOM$_1$=ER/LV\cite{Rahim2021}, and maximum static ER normalised by IL: FOM$_{2}$=ER/IL. ER needs to be maximised to allow multi-level modulation formats such as PAM-4, because multiple bits must be encoded without impacting the BER at the receiver\cite{Rahim2021}.

Table \ref{tab1:Modulators} reports amplitude modulators (AMs) in silicon photonics (SiPh)\cite{Reed2010, Rahim2021}, based on Si\cite{Li2024, Han2023, Zhou2019, Sakib2021, Yuan2024, Chan2024}, Ge\cite{Srinivasan2016, Liu2022, Hu2022}, III-V\cite{Shahin2019}, lithium niobate (LN\cite{He2019}), or plasmonic structures\cite{Eppenberger2023}. In Si devices, Mach-Zehnder modulators (MZMs)\cite{Han2023, Zhou2019} and ring modulators (RMs)\cite{Sakib2021} exploit the plasma dispersion effect\cite{Soref1987}, in which a variation of carrier density results in a change of the refractive index\cite{Soref1987}. Power consumption in MZMs is typically$\sim$pJ/bit\cite{Thomson2016a}, due to low modulation efficiency ($V_{\pi}L\sim$1.2Vcm\cite{Zhou2019}) and large footprint of the phase shifters (several mm\cite{Zhou2019}) for carrier depletion type modulators\cite{Zhou2019}, and due to high losses (5dB/mm\cite{Gardes2018}) for carrier accumulation type modulators\cite{Gardes2018}. RMs are better than MZMs in terms of footprint ($\sim$400$\mu$m$^2$\cite{Chan2024}) and energy consumption (5.3fJ/bit\cite{Sakib2021}), but they are strongly temperature-sensitive because they have a narrow optical bandwidth (BW, sub-nm\cite{Sakib2021}) which shifts by tens of picometers per degree (e.g. 77.5pm/K in Ref.\cite{Janz2024}).
\begin{table*}[t]
    \centering
    \footnotesize 
    \begin{adjustbox}{max width=\linewidth}
    \begin{tabular*}{\linewidth}{@{\extracolsep{\fill}} l l l c c l l l l l l }
        \toprule
        Ref. & Material & IL [dB] & \multicolumn{2}{c}{ER [dB]} & Band & $\frac{ER}{LV}$ [$\frac{dB}{\mu mV}$] & $\frac{CV_{pp}^2}{4}$ [$\frac{fJ}{bit}$] & Active area [$\mu$m] & f$_{3dB}$ [GHz] & NRZ rate [Gbit/s] \\
        \cmidrule(lr){4-5}
         &  &  & Static & Dynamic &  &  &  &  &  &  \\
        \midrule
        \cite{Han2023} & Si MZM          & 6.8  & -    & 2.15   & C & -      & -    & 124 $\times$ 0.835 & 110  & 112 \\
        \cite{Sakib2021} & Si RM           & -    & -    & 3.8    & O & -      & 5    & -                & 77   & 128 \\
        \cite{Chan2024}  & Si RM           & 0.9  & 16   & -      & O & 0.053  & 6.3  & 72               & 49   & 180 \\
        \cite{Srinivasan2016} & Ge         & 4.9  & 4.6  & 3.8    & L & 0.05   & 12.8 & 40 $\times$ 0.6  & $>$50 & 56  \\
        \cite{Liu2022}  & Ge              & 5.7  & 14.1 & 5      & L & 0.1    & 6.3  & 40 $\times$ 0.6  & $>$67 & $>$80 \\
        \cite{Hu2022}   & Ge              & 6.9  & 12.6 & 2.2    & L & 0.04   & -    & 25 $\times$ 1    & $>$67 & 110 \\
        \cite{Shahin2019} & III-V         & 5    & 15   & -      & C & 0.037  & -    & 200              & 33   & 80  \\
        \cite{He2019}   & LN              & 2.5  & 40   & 5      & C & 0.001  & 170  & 5000             & $>$70 & 100 \\
        \cite{Eppenberger2023} & Plasmonic RM & 1.5 & 5.2 & -      & C & 0.13   & 29   & 10 $\times$ 0.1  & 176  & 220 \\
        \cite{Giambra2019} & DSLG EAM     & 20   & 3    & 1.3    & C & 0.0025 & -    & 120 $\times$ 0.65& 29   & 50  \\
        \cite{Agarwal2021} & DSLG EAM     & 7.8  & 4.4  & 5.2    & C & 0.037  & 160  & 60 $\times$ 0.45 & 39   & 40  \\
        *20 nm t$_{Ox}$   & DSLG EAM     & 1    & 3    & 1      & C & 0.037  & 26   & 20 $\times$ 0.65 & 17   & 20  \\
        *40 nm t$_{Ox}$   & DSLG EAM     & 0.9  & 4    & 1.2    & C & 0.01   & 58   & 40 $\times$ 0.55 & 67   & $>$80\\
        \bottomrule
    \end{tabular*}
    \end{adjustbox}
    \caption{Performance comparison between AMs based on Si, III-V (InGaAsP), LN and graphene. *This work.}
    \label{tab1:Modulators}
\end{table*}

Si MZM and RM achieved 112\cite{Han2023} and 180Gbit/s\cite{Chan2024} NRZ data rates, respectively. SiGe EAMs based on the Franz-Keldysh\cite{Keldysh1957} or quantum-confined Stark\cite{Weiner1987} effects were reported up to 110Gbit/s\cite{Hu2022}. SiGe EAMs have smaller footprint ($\sim$400$\mu$m$^2$\cite{Srinivasan2016}), lower energy consumption ($<$150fJ/bit\cite{Srinivasan2016}) than MZMs, and wider optical BW than RM ($>$20nm\cite{Srinivasan2016}). However, they cannot operate in the O-band (1260-1360nm) due to the $E_g\sim$0.7eV direct bandgap in Ge\cite{Kittel1996}, limiting the operating wavelength to $\lambda=hc/E_g\sim$1600nm. LN on Si modulators achieved 100Gbit/s\cite{He2019}, but $V_{\pi}L\sim$2.2Vcm\cite{He2019} is such that several mm long phase shifters\cite{He2019} in a Mach-Zehnder configuration are required\cite{He2019}, hence footprint and energy consumption are large (170fJ/bit\cite{He2019}). III-V on Si AMs were reported with FOM$_1$=ER/IL$\sim$ 3\cite{Shahin2019} and NRZ data rates up to 80Gbit/s\cite{Shahin2019}. Plasmonic ring modulators achieved 220Gbit/s NRZ\cite{Eppenberger2023} and ultra-miniaturisation of the active area$\sim$10$\times$0.1$\mu$m\cite{Eppenberger2023}. However, they suffer from intrinsically large IL (0.4dB/$\mu$m) due to the presence of metal near the waveguide (WG)\cite{Eppenberger2023}. 

SiPh alone, or the integration of Ge, III-V, LiNbO$_3$, or plasmonic structures, do not fully satisfy the joint metrics of high data rate ($>$50GBaud per lane\cite{EthernetAlliance2024}), low footprint\cite{Reed2010} ($<$1mm), broadband (from O to L band, i.e. 1260-1625nm) operation, and low energy consumption per bit ($<$ 100fJ/bit\cite{Reed2010}). It is thus necessary to explore alternative materials that can provide broad wavelength operation in datacom bands (1260-1625nm), and with a CMOS-compatible manufacturing process.

Single Layer Graphene (SLG) has ambipolar field effect\cite{Geim} and broadband (500nm-10$\mu$m\cite{Li2008, Nair2008}) gate-variable optical conductivity\cite{Wang2008}, giving rise to electro-optic effects exploitable in datacom bands\cite{Romagnoli2018}. SLG is excellent for photonic applications\cite{Romagnoli2018, Bonaccorso2010, Koppens2014}, in particular for on-chip modulation\cite{Liu2011, Liu2012, Phare2015, Sorianello2018, Giambra2019, Agarwal2021, Watson2023}, detection\cite{Muench2019c, Schulera, Miseikis2019, Marconi2021}, saturable absorption\cite{Sun2010}, and photomixing\cite{Montanaro2016, Montanaro2021, Montanaro2023, Montanaro2024}, with disruptive potential for next-generation optical communications\cite{Romagnoli2018}. 

The SLG field-effect charge carrier mobility at room temperature (RT) ($\mu>$100,000cm$^2$/Vs in suspended SLG)\cite{Banszerus2015, Purdie2018, DeFazio2019, Banszerus2016, Banszerus2019}, is 20 times higher than Ge (3,900cm$^2$/Vs\cite{Prince1953}) and 10 times higher than Ga$_x$In$_{1-x}$As (12,000cm$^2$/Vs\cite{Adachi1992}). Because of the Van der Waals nature of the interaction between SLG and substrate, the lattice matching condition is relaxed\cite{Koma1999}, hence SLG can be transferred on any substrate\cite{Koma1999, Suk2011}. Ge and III-V films are grown by epitaxy on Si, with$>$4\% lattice mismatch\cite{Ye2014, Kunert2018}. This leads to a strain energy larger than that required to break a bond\cite{Gabrys2018}, causing the formation of defects during epitaxy\cite{Ye2014, Kunert2018}, which hinders large size (e.g. 8" wafers) fabrication\cite{Liu2020}.

Double-SLG (DSLG) EAMs on SiPh platforms\cite{Liu2011, Liu2012, Phare2015, Sorianello2018, Giambra2019, Agarwal2021, Watson2023} can have large (0.037dB/V$\mu$m\cite{Agarwal2021}) FOM$_2$ for a given DC bias and V$_{pp}$, because $\mu$ affects the slope of the absorption as a function of voltage\cite{Sorianello2020}. The higher the $\mu$, the higher ER/V. Additionally, ER is large ($>$0.1dB/$\mu$m\cite{Agarwal2021}) if the SiPh platform is engineered for SLG integration\cite{Agarwal2021}, enabling short ($L<$100$\mu$m\cite{Agarwal2021}) devices and low ($<$1dB\cite{Agarwal2021}) IL. This reduces the energy consumption per bit. DSLG EAMs also counteract chromatic dispersion\cite{Sorianello2017, Sorianello2019}, because they show a positive linear chirp associated to electric gating responsible for the absorption variations\cite{Sorianello2017, Sorianello2019}, which can be exploited for chromatic dispersion compensation\cite{Sorianello2017, Sorianello2019}. Thus, DSLG EAMs are promising for DSP-free NRZ-based transceivers for$<$500m optical interconnects\cite{Verbist2018}, and even$>$500m, due to their high EO-BW (39GHz\cite{Agarwal2021}).

Here, we report wafer-scalable DSLG EAMs on Si-on-insulator (SOI) WGs\cite{Littlejohns2020} using SLG grown by chemical vapour deposition (CVD). We get FOM$_1\sim$0.037dB/V$\mu$m, compact footprint (active area$\sim$13$\mu$m$^2$), low energy consumption ($\sim$26fJ/bit) and 20Gbit/s NRZ data rate, as well as modulators with FOM$_1\sim$0.01dB/V$\mu$m, EO-BW=67GHz, and NRZ data rate$>$80Gbit/s, with$\sim$22$\mu$m$^2$ active area and$\sim$58fJ/bit. IL$<$0.05dB/$\mu$m, with SLG$\sim$10nm from the active sections, thus maximising the interaction with the evanescent field, and reaching average static ER$\sim$0.12dB/$\mu$m. We report Fermi level, E$_F$, tuning$\sim$0.64 eV by solid-state gating. Table \ref{tab1:Modulators} summarises the performance of our modulators compared with SiGe, LN, III-V and other SLG AMs. We report modulation efficiency$\sim$0.037dB/V$\mu$m, on par with SLG and hBN exfoliated flakes on a transverse-magnetic polarized WG\cite{Agarwal2021}, and$\sim$16 times larger than other CVD SLG\cite{Giambra2019}. This is$\sim$40 times better than LN on Si\cite{He2019}, and on par with III-V on Si\cite{Shahin2019}. Our devices are half the size of previous SLG\cite{Agarwal2021}, and SiGe\cite{Liu2022} ones,$\sim$10 times shorter than III-V on Si\cite{Shahin2019}, and$\sim$250 times shorter than LN\cite{He2019}. This translates in an energy consumption ($\sim$26fJ/bit), $\sim$6.5 times smaller than MZM based on Si\cite{Gardes2018} or LN\cite{He2019}. We reach 3dB EO-BW (f$_{3 dB}=$67GHz) almost double that of previous SLG modulators\cite{Phare2015, Giambra2019, Agarwal2021}, with NRZ data rate$\sim$80 Gbit/s, $\sim$1.6 times larger\cite{Giambra2019}, on par with SiGe EAMs\cite{Srinivasan2016, Liu2022, Hu2022}, but with operation in both O and C bands. Thus, our modulators are useful for both short-reach IMDD and long-reach coherent communications; 1.6Tbit/s is possible by designing a photonic integrated circuit with 20 lanes, each with a 80Gbit/s NRZ modulator.
\begin{figure*}
\centerline{\includegraphics[width=180mm]{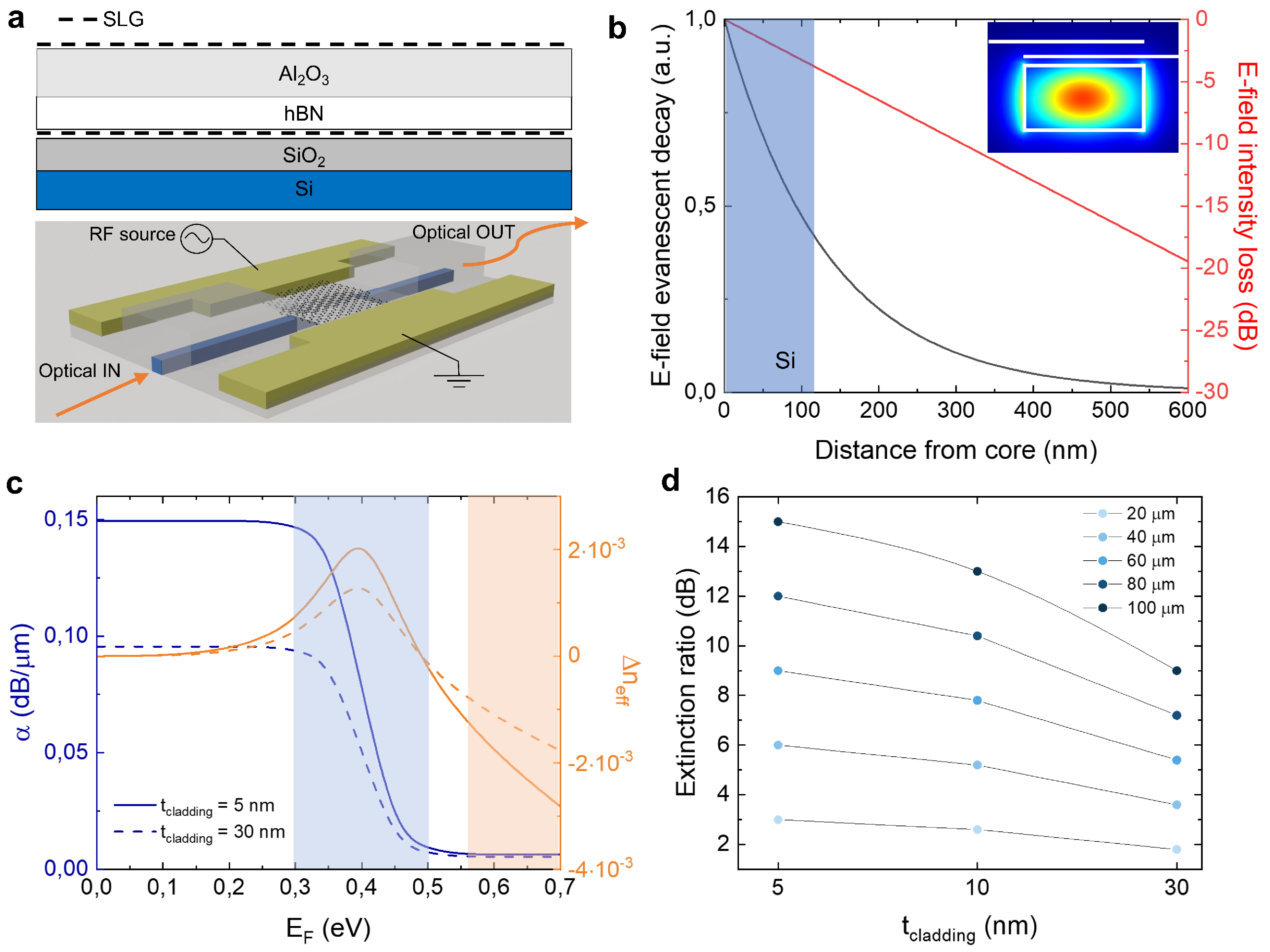}}
\caption{\textbf{a} Cross-section of DSLG modulator, showing metal pads, structured SiO$_2$ cladding, DSLG capacitor and device operating principle. The RF signal, applied to the DSLG capacitor metal pads, modulates the intensity of a laser coupled to the WG. The modulated laser is then transmitted out of the device. \textbf{b} Linear and logarithmic plots of electric field vertical decay as a function of distance from WG core. The inset shows the electric field in the WG (dashed rectangle), the evanescent field, and the DSLG capacitor (white lines are SLG layers, embedded in hBN/Al$_2$O$_3$ dielectric). The evanescent field, located outside the Si core, interacts with the DSLG stack, enabling electro-optic effects. The DSLG should be as close as possible to the evanescent field, meaning that the oxide cladding between Si and DSLG should be minimized. \textbf{c} Simulated absorption and change in WG refractive index as a function of E$_F$ for $\lambda=$1550nm and two oxide thicknesses. The blue region is the operating point for electro-absorption, while the orange one is that for electro-refraction. The DSLG capacitor has a 20nm gate oxide and 0.65$\mu$m gated SLG length. \textbf{d} Simulated maximum ER as vs oxide thickness for lengths 20, 40, 60, 80, 100$\mu$m, from \textbf{c}.}
\label{Fig1:EvanescentCoupling}
\end{figure*}
\begin{figure*}
\centerline{\includegraphics[width=180mm]{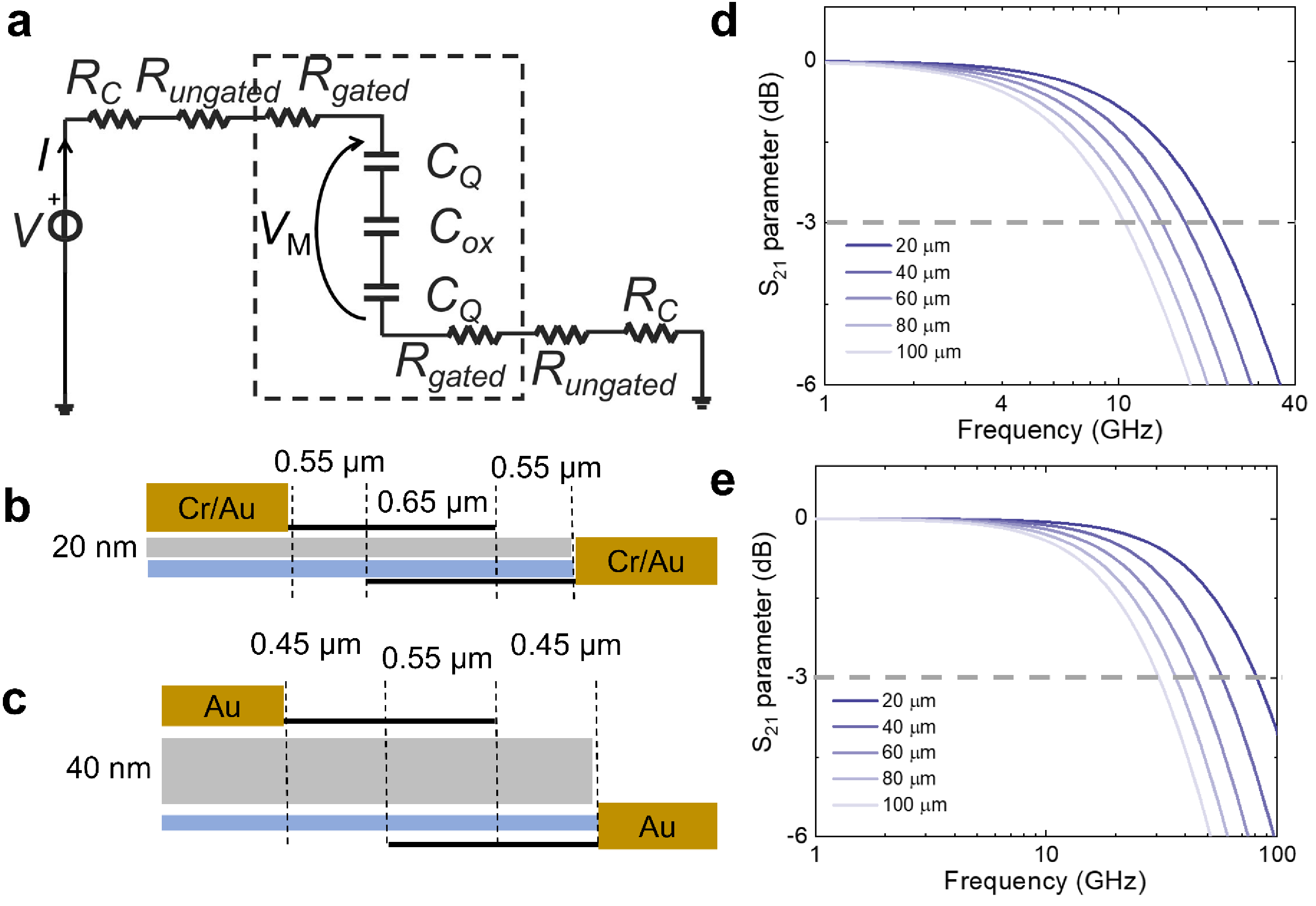}}
\caption{\textbf{a} Lumped element model of DSLG modulator, with R$_C$, R$_{ungated}$, R$_{gated}$ and C$_Q$, C$_{ox}$ components. \textbf{b} Cross-section of DSLG modulator design with 20nm-thick gate oxide, $\epsilon_{hBN-Al2O3}$= 6.35, 0.65$\mu$m gated SLG length, 0.55$\mu$m ungated SLG length, R$_C$=1k$\Omega\mu$m; and \textbf{c} with 40 nm-thick gate oxide, $\epsilon_{hBN-Al2O3}$= 6.8, 0.55$\mu$m gated SLG length, 0.45$\mu$m ungated SLG length, R$_C$=600$\Omega\mu$m. \textbf{d} Simulated frequency response of 20nm-thick and \textbf{e} 40nm-thick gate oxide designs. Both use our average experimental $\mu\sim$8,000cm$^2$/Vs, which corresponds to $\tau\sim$350fs. Different contact resistances are used to model Cr/Au and Au contacts. The design in \textbf{c, e} targets broadband RF operation. For a 40$\mu$m-long modulator, the design in \textbf{c, e} shows 70GHz EO-BW, while \textbf{b, d} give 24GHz EO-BW.}
\label{Fig2:BW}
\end{figure*}
\section{Device design}
The SLG density of states (DOS) near the Dirac point can be written as\cite{Wallace1947}:
\begin{equation}\label{EqDOS}
  DOS(E) = \frac{2|E_F|}{\pi(\hbar v_F)^2} ,
\end{equation}
with $\hbar$ the reduced Planck's constant, and $v_F\sim9.5\times10^5$m/s\cite{Romagnoli2018} the Fermi velocity. Eq.\ref{EqDOS} gives$\sim$10$^{15}$eV$^{-1}$m$^{-2}$ for E$_F$=$\pm$50meV. The SLG DOS was reported to be$\sim$10$^{17}$eV$^{-1}$m$^{-2}$ for E$_F$=$\pm$50meV due to substrate-induced disorder\cite{Droscher2012, Martin2008}, one order of magnitude lower than 2d electron gases with parabolic dispersion, e.g. GaAs, with $DOS=m^*/\pi\hbar^2\sim$10$^{18}$eV$^{-1}$m$^{-2}$\cite{Sze2006}, using $m^*=0.063m_e$\cite{Sze2006}. As a result, the $E_F$ shift as a function of charge carrier density, $n$, in SLG is larger than in GaAs. For $n=10^{16}$m$^{-2}$, $dE_F/dn=\hbar v_F\sqrt{\pi}/2\sqrt{n}\sim$60meV in SLG\cite{CastroNeto2009}. For the same $n$, $dE_F/dn=\pi\hbar^2/m^*\sim$10meV in GaAs\cite{Sze2006}. 

The large $dE_F/dn$ of SLG can be exploited in the DSLG EAMs, based on a SLG-dielectric-SLG capacitor coupled to a WG, Fig.\ref{Fig1:EvanescentCoupling}\textbf{a}. In DSLG EAMs, the application of a voltage between two SLGs creates an electric field perpendicular to the WG, and changes E$_F$ of both SLGs. This can be used to control the optical conductivity $\sigma(\omega)$ of both SLGs electrostatically\cite{Wang2008, Liu2012}, therefore the complex effective index $n=n_{eff}+i\kappa$ of the electromagnetic mode propagating in the WG\cite{Sorianello2016}, where the real part accounts for the mode's phase\cite{Hect2017} and the imaginary one for the loss\cite{Hect2017}. The induced effective refractive index change $\Delta n_{eff}$ leads to phase modulation $\Delta\phi=2\pi L/\lambda_0\Delta n_{eff}$\cite{ReedBook2004}, while the variation of the extinction coefficient $\Delta \kappa$ is related to a variation of the linear absorption coefficient $\Delta\alpha=4\pi/\lambda_0\Delta\kappa$\cite{ReedBook2004} along the direction of propagation. The latter is exploited to generate intensity modulation\cite{Romagnoli2018}.

The DSLG EAM modulation efficiency is affected by the overlap between SLGs and modal field profile, specifically by the evanescent field, decaying exponentially within the cladding as per Beer-Lambert law I=I$_0$exp(-$\Gamma$$\alpha$L)\cite{Beer1852}, as expressed by the confinement factor\cite{Robinson2008}:
\begin{equation}
  \Gamma = \frac{n_g}{n_{cl}}\times \frac{\iint_{cl} \epsilon_{cl}|E|^2 dxdy}{\iint_{\infty} \epsilon |E|^2 dxdy},
\end{equation}
where the first term is the ratio between group refractive index and real part of the cladding refractive index, and the second represents the normalized electric field energy density in the cladding\cite{Robinson2008}. Fig.\ref{Fig1:EvanescentCoupling}\textbf{a} is a DSLG EAM, comprising a 220nm SOI WG with a structured cladding and a SLG/hBN/Al$_2$O$_3$/SLG capacitor integrated on the active area, identified as the region with thin ($<$30nm) cladding. Fig.\ref{Fig1:EvanescentCoupling}\textbf{a} also shows the optical input and intensity-modulated output from the DSLG structure modulated by a RF signal. We simulate TE-polarized single-mode SOI channel WGs at 1.55$\mu$m using the finite difference element (FDE) MODE solver in Lumerical\cite{Lumerical}. The evanescent field decays within the cladding at 0.03dB/nm, Fig.\ref{Fig1:EvanescentCoupling}\textbf{b}. At the interface between the Si core and the SiO$_2$ cladding, the electric field intensity drops by$\sim$3.5dB. As the cladding thickness increases, the SLG-light interaction reduces. Hence, we investigate the dependence of $\Delta\kappa$ and $\Delta n_{eff}$ on cladding thickness. Refs.\cite{Giambra2019, Giambra2021} used 40\cite{Giambra2021} and 10nm\cite{Giambra2019} thick claddings. Ref.\cite{Giambra2019} used numerical simulations to predict ER$\sim$0.13dB/$\mu$m between 0 and 20V, with a 10nm thick cladding, but measured 0.03dB/$\mu$m\cite{Giambra2019}. Here, we calculate $\alpha$ for different cladding thicknesses 30, 10, 5nm, and the resulting maximum ER for different lengths (from 20 to 100$\mu$m) Fig.\ref{Fig1:EvanescentCoupling}\textbf{c},\textbf{d}, estimating 0.15, 0.13, 0.09dB/$\mu$m for 5, 10, 30nm, respectively. $P_{min}$ and $P_{max}$ are calculated at the SLG maximum and minimum absorption, which correspond to tuning $E_F$ from 0 to 0.5eV.

We model SLG as a surface conductivity derived from Kubo’s formula\cite{Kubo1957,Gusynin2007}:
\begin{equation}
\begin{split}
    \sigma(&\omega, \mu_C, \Gamma, T) =  \frac{ie^2}{\pi \hbar^2 (\omega - i2\Gamma)} \\
    &\int_{0}^{\infty}E\biggl[\frac{\partial f_d(E)}{\partial E}-\frac{\partial f_d(-E)}{\partial E}\biggl]dE \\
    & - \frac{ie^2 (\omega - i2\Gamma)}{\pi \hbar^2} \int_{0}^{\infty}\frac{f_d(-E)-f_d(E)}{(\omega - 12\Gamma)^2 - 4(E/\hbar)^2}dE ,
\end{split}
\end{equation}
where $\omega$ is the angular frequency, $\Gamma$ is the scattering rate, $f_d(E)=1/[e^{(E-E_F)/(k_B T)}+1]$ is the Fermi-Dirac distribution, and T the temperature\cite{Hanson2008DyadicGraphene}. The first term relates to intraband transitions\cite{Hanson2008DyadicGraphene, Falkovsky2007}, while the second to interband ones\cite{Hanson2008DyadicGraphene, Falkovsky2007}. $\Gamma$ is related to the transport relaxation time $\tau=\hbar/\Gamma$\cite{Sorianello2020}. We use $\tau$=350fs, corresponding to our experimental $\mu\sim$8,000cm$^2$/Vs, using $\mu\sim e\tau v^2_F/E_F$ for $E_F>>k_{B}T$\cite{Romagnoli2018}, with $v_F\sim9.5\times10^5$m/s\cite{Romagnoli2018}. When $2E_F>\hbar c/\lambda$, SLG enters the Pauli blocking regime\cite{Romagnoli2018}, and interband transitions are blocked\cite{Pisana2007}, while for $2E_F<\hbar c/\lambda$ interband transitions are allowed, hence electro-absorption is maximum. In the telecom C-band the wavelength is 1550nm, which corresponds to $\hbar c/\lambda$=0.8eV. Hence, SLG enters transparency for $E_F>$0.4eV. EAMs work in the absorptive regime, while PMs in the transparent one\cite{Watson2023}. These regions are illustrated in Fig.\ref{Fig1:EvanescentCoupling}\textbf{c} as blue (absorptive) and orange (transparent). In the absorptive regime, we calculate $\alpha$=0.09dB/$\mu$m for $t_{cladding}$=30nm and $\alpha$=0.15dB/$\mu$m for 5nm. This corresponds to a$\sim$40$\%$ increase in SLG absorption. It is thus key to thin down SiO$_2<$30nm, to optimise ER and modulator FOMs.

The EO-BW depends on the mechanism responsible for the EO conversion. Since in DSLG modulators this occurs by charge carriers accumulation in the SLGs forming a capacitor, EO-BW depends on the electrical BW of such capacitor\cite{Sorianello2015}, i.e. the frequency at which the electrical power transferred from generator to load decreases by 3dB\cite{Pozar}. Hence, we model the modulator as idealized resistances and capacitances (lumped element\cite{Bahl2003}), to calculate the voltage drop in the capacitor leading to the drop in optical power transmitted by the modulator. Lumped element modelling is allowed because the equivalent RC circuit length ($<$100$\mu$m) is smaller than the applied RF wavelength (1-3mm). 

With reference to Fig.\ref{Fig2:BW}\textbf{a}, $C_{tot}=(C_{Q}C_{ox})/2(C_{Q}C_{ox})$ is the series of two capacitances. The first is the geometrical capacitance $C_{ox}=(\epsilon_0\epsilon_r A)/d$, where $\epsilon_0$ is the vacuum, $\epsilon_r$ is the permittivity of the dielectric spacer between SLGs, A and d are the area and thickness of dielectric spacer respectively. The second is the quantum capacitance, associated to the 2d electron gas\cite{Das2008}, $C_Q=(2e^2\sqrt{n_{tot}})/(\hbar v_F\sqrt\pi)$. $R_T=R_S+2(R_{C}+R_{ungated}+R_{gated})$ is the series of the generator impedance (R$_S$=50$\Omega$), the contact resistance $R_C$ and the SLG sheet resistances $R_{gated}$ and $R_{ungated}$. Gated and ungated SLG sections have different sheet resistance, $\tau$, and E$_F$, because they operate at the quadrature point ($E_F\sim$0.4eV), while ungated sections remain at the original $E_F\sim$0.2eV. $R_{ungated}$ corresponds to SLG sections not part of the capacitor, with fixed carrier concentration, while $R_{gated}$ refers to gate-tunable SLG sections. Fig.\ref{Fig2:BW}\textbf{b} shows the simulated S$_{21}$ for a DSLG with 650nm gated section, 550nm ungated sections, 20nm thick hBN/Al$_2$O$_3$ dielectric and a contact resistance R$_C$=1k$\Omega$$\mu$m, while Fig.\ref{Fig2:BW}\textbf{c} shows a design optimized for speed, with 550nm gated section, 450nm ungated sections, 40nm thick hBN/Al$_2$O$_3$ dielectric and R$_{C}W$=600$\Omega$$\mu$m. We fix $E_F$=0.4eV for gated SLG and 0.2eV for ungated. We then take $\tau_{gated}$=350fs ($R_{gated}\sim$60$\Omega/\square$ and $\mu\sim$8,000cm$^2$/Vs). We use these to calculate $\tau_{ungated}\sim$210fs ($R_{ungated}\sim$170$\Omega/\square$).

By modelling how SLG contributes to the total resistance and capacitance of the circuit, we calculate the circuit impedance and analyse its frequency response. The impedance is given by $Z_C=1/(i\omega C_{tot})$ and the voltage drop in the capacitor is $V_C(\omega)=I(\omega)Z_C$. The total impedance $Z_T=R_T+Z_C$ also includes the total resistance of the circuit, leading to a total driving voltage $V_T=I(\omega)Z_T=I(\omega)(R_T+Z_C)$. From this, we calculate the frequency response of the modulator as $V_M={V_C(\omega)}/{V_T(\omega)}={Z_C}/{Z_T}={1}/{1+i\omega R_T C_T}$. We then calculate the drop in transmitted power as $P_{out}/P_{in}=20log⁡(|{V_C}/{V_T}|)$ and extrapolate the 3dB cut-off for different modulator designs (Fig.\ref{Fig2:BW}\textbf{b}, \textbf{c}) and lengths, as in Fig.\ref{Fig2:BW}\textbf{d},\textbf{e}. $f_{3dB}$ increases as we reduce the length of the modulator for fixed contact resistance and $\mu$. The design in Fig.\ref{Fig2:BW}\textbf{b} achieves a maximum EO-BW$\sim$28GHz for L=20$\mu$m (Fig.\ref{Fig2:BW}\textbf{d}), mainly limited by the high ($>$1k$\Omega$) contact resistance of Cr/Au contacts\cite{Nagashio2009}, and by the resistance contribution of gated and ungated sections. The design optimized for high speed in Fig.\ref{Fig2:BW}\textbf{c} achieves EO-BW$\sim$80GHz (Fig.\ref{Fig2:BW}\textbf{e}), enabled by using a thicker dielectric, which reduces capacitance, and by a lower contact resistance, thanks to the Au contact\cite{Sundaram2011, Cusati2017}.
\begin{figure*}
\centerline{\includegraphics[width=180mm]{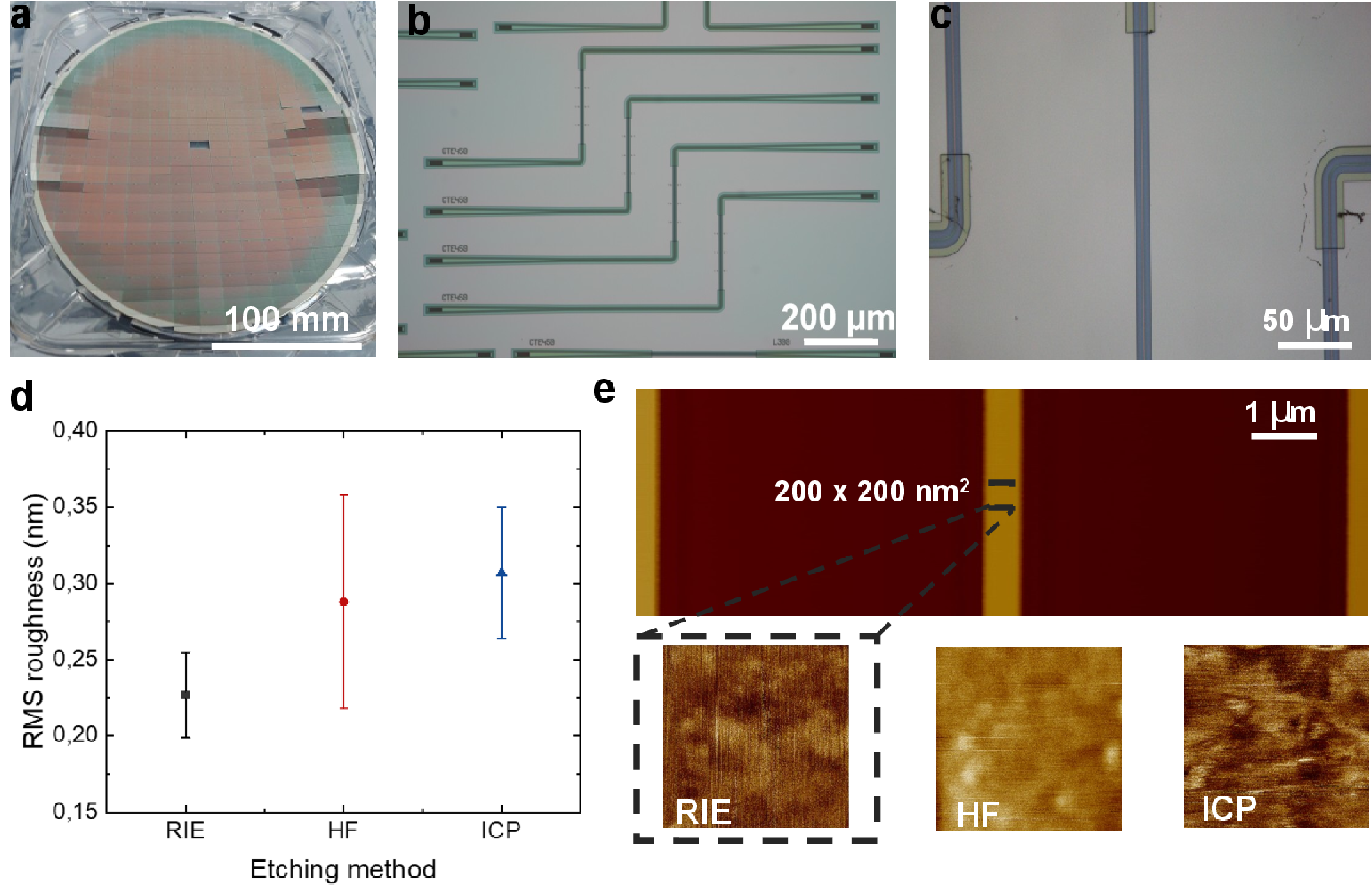}}
\caption{\textbf{a} 200mm SOI wafer containing passive integrated photonics components such as WGs, grating couplers, and interferometers, manufactured with the dedicated structured cladding process for SLG integration. \textbf{b} Optical image of S-bend SOI WGs before SLG transfer. \textbf{c} SOI WGs after polycrystalline, continuous SLG film wet transfer. While SLG might break on the passive region, it is continuous on the active area, where the device is fabricated. \textbf{d} RMS roughness of SiO$_2$ surface after different etching methods, extrapolated from \textbf{e} 200$\times$200nm$^2$ AFM topography images taken on the WG active area.}
\label{fig3:SOI-Wafer}
\end{figure*}
\begin{figure*}
\centerline{\includegraphics[width=180mm]{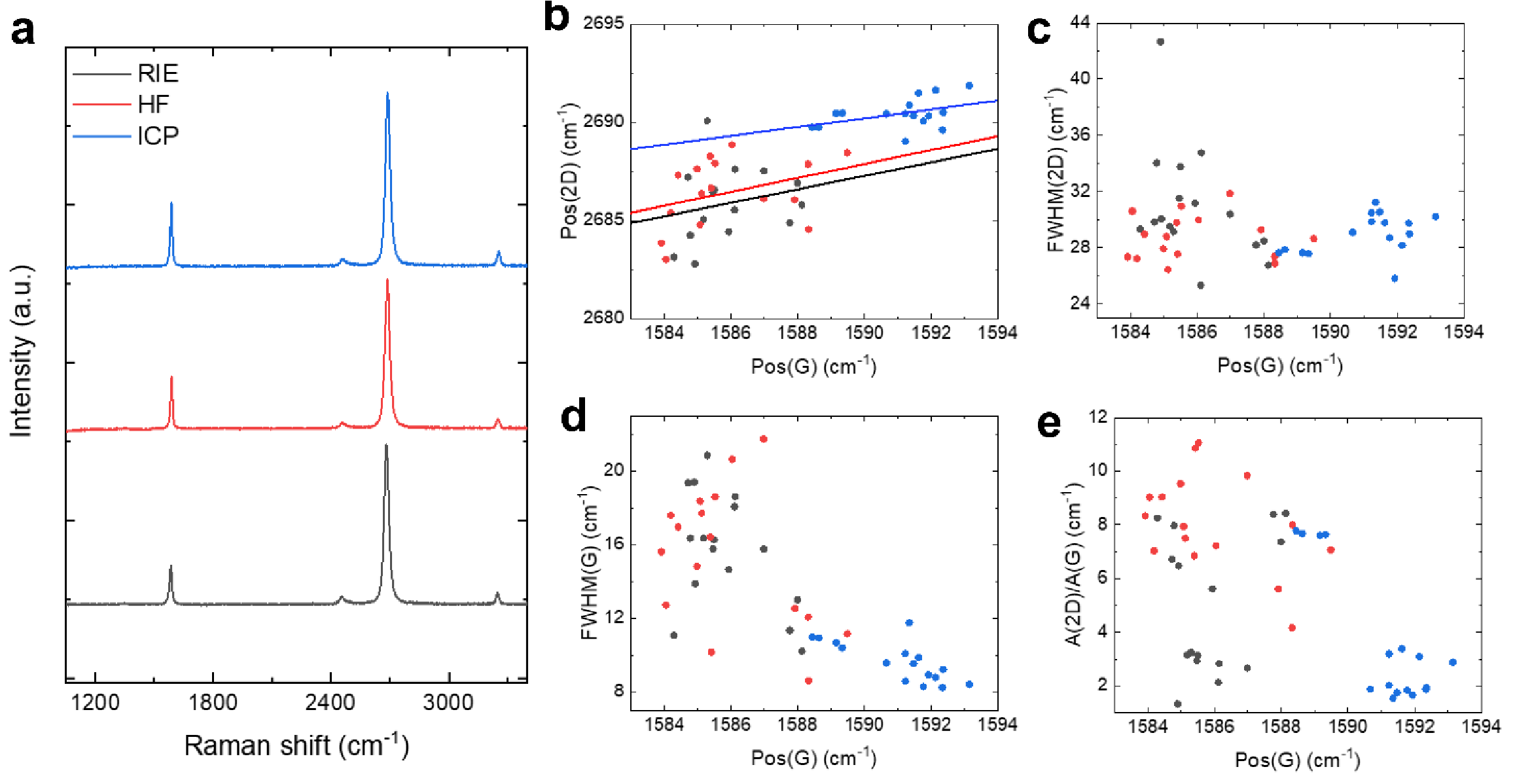}}
\caption{\textbf{a} Raman spectra of SLG transferred on SOI after RIE, HF, and ICP etching of SiO$_2$ cladding. \textbf{b-e} Correlation plots collected from 48 measurements on RIE, HF and ICP etched substrates. \textbf{b} Pos(2D) as a function of Pos(G). \textbf{c} FWHM(2D) as a function of Pos(G). \textbf{d} FWHM(G) as a function of Pos(G). \textbf{e} A(2D)/A(G) as a function of Pos(G).}
\label{fig4r:Raman}
\end{figure*}
\begin{table*}[t]
    \centering
    \begin{tabular*}{\linewidth}{@{\extracolsep{\fill}}c c c c c}
    \toprule 
     Samples & SLG on Cu & SLG on SOI (RIE-etched) & SLG on SOI (ICP-etched) & SLG on SOI (HF-etched) \\ 
    \midrule 
    Pos(G) (cm$^{-1}$) & 1584 $\pm$ 1 & 1586 $\pm$ 1 & 1590 $\pm$ 3 & 1586 $\pm$ 2 \\
    FWHM(G) (cm$^{-1}$) & 9 $\pm$ 1 & 15 $\pm$ 2 & 12 $\pm$ 6 & 14 $\pm$ 3 \\
    Pos(2D) (cm$^{-1}$) & 2698 $\pm$ 1 & 2686 $\pm$ 3 & 2690 $\pm$ 1 & 2687 $\pm$ 1 \\ 
    FWHM(2D) (cm$^{-1}$) & 24 $\pm$ 1 & 28 $\pm$ 7 & 29 $\pm$ 1 & 29 $\pm$ 1 \\
    A(2D)/A(G) & 1.9 $\pm$ 0.4 & 3.3 $\pm$ 2.2 & 3 $\pm$ 2 & 3.8 $\pm$ 1 \\
    I(2D)/I(G) & 2.6 $\pm$ 0.3 & 2.1 $\pm$ 0.9 & 1.5 $\pm$ 1.1 & 3.1 $\pm$ 0.8 \\
    I(D)/I(G) & N.A & 0.04 $\pm$ 0.30 & 0.09 $\pm$ 0.10 & 0.02 $\pm$ 0.01 \\
    E$_F$ (meV) & 291 $\pm$ 200 & 234 $\pm$ 99 & 360 $\pm$ 145 & 187 $\pm$ 57 \\
    Doping type & p & p & p & p \\
    n ($\times$ 10$^{12}$ cm$^{-2}$) & 10 $\pm$ 11 & 13.7 $\pm$ 13.2 & 20 $\pm$ 19 & 9.2 $\pm$ 5.6 \\
    Uniaxial strain (\%) & 0.14 $\pm$ 0.28 & -0.05 $\pm$ 0.09 & 0.05 $\pm$ 0.21 & 0.07 $\pm$ 0.21 \\
    Biaxial strain (\%) & 0.05 $\pm$ 0.10 & -0.02 $\pm$ 0.03 & 0.02 $\pm$ 0.08 & 0.03 $\pm$ 0.08 \\
    n$_D$ ($\times$ 10$^{10}$ cm$^{-2}$) & N.A. & 0.85 $\pm$ 0.70 & 3.6 $\pm$ 3.9 & 0.68 $\pm$ 0.21 \\
    \bottomrule 
    \end{tabular*}
    \caption{Raman peaks analysis and corresponding E$_F$, doping type, $n$, strain, n$_D$ and related uncertainties, for SLG transferred on 3 SOI platforms, with different SiO$_2$ etching methods: RIE, HF and ICP.}
    \label{tab2:RamanEtchTest}
\end{table*}

\section{Graphene integration}
To enable wide adoption of SLG in integrated photonics, it is essential to devise a scalable, reproducible, and CMOS-compatible fabrication flow. Here, we use the 200mm SOI fabrication process of Ref.\cite{Littlejohns2020} to produce the chips for the integration of SLG, and hBN grown by CVD. Wafer-scale SLG integrated photonic components currently rely on the transfer from growth substrate to target photonics platform\cite{Giambra2021}. Although the transfer does not suffer from lattice mismatch with the substrate\cite{Koma1999}, unlike III-V and Ge-Si platforms\cite{Ye2014,Kunert2018}, it can introduce wrinkles\cite{Suk2011} and defects\cite{Suk2011}, as well as unwanted optical losses\cite{Giambra2019, Giambra2021}. One approach is to grow matrices of single crystal SLGs and transfer them onto the wafer after alignment to photonic circuits\cite{Giambra2021, Miseikis2021}. This exploits metallic nucleation seeds for the growth of SLG crystals\cite{Miseikis2017} with a pitch matching the photonic circuits on the wafer (dimensions up to 350 $\mu$m\cite{Miseikis2017, Giambra2021}). The position of the crystal matches that of the photonic circuit, so that SLG covers only specific regions of interest. However, this method has disadvantages. 1) The scalability is non-trivial\cite{Miseikis2021}, and there are more steps compared to growth and transfer of wafer scale SLG, as the growth substrate needs to be pre-processed for nucleation seeds\cite{Wu2011}. 2) The delamination of single crystals leaves parasitic metallic particles below the crystals\cite{Miseikis2017}, which might lead to optical losses. 3) Multi-layer graphene is typically observed at the centre of the crystals\cite{Miseikis2017}, reducing the usable SLG area, and creating alignment issues when stamping on the optical WGs\cite{Giambra2021}, and for the DSLG stack fabrication\cite{Giambra2021}. Another approach is to transfer SLG films from Cu on the whole target substrate\cite{Li2009a, Suk2011}, removing SLG later by etching. In SiPh technologies, this is typically done after planarizing the oxide by chemical-mechanical polishing (CMP)\cite{Absil2015}. In both cases, SLG-integrated photonic devices have both active and passive regions exposed to the transfer process, which might induce losses.

Here, we combine CMP and dry etching to planarize the oxide cladding and selectively lower it, so that only the active sections of the devices are exposed to SLG transfer, while the passive sections are protected by a 1$\mu$m thick cladding. Passive blocks are optically isolated from materials and contaminants, while active ones are integrated with the material. We first planarize the SiO$_2$ surface of a 200mm SOI wafer (Fig.\ref{fig3:SOI-Wafer}\textbf{a}) by CMP down to 1$\mu$m. Then, we etch SiO$_2$ leaving$<$30nm, except on top of the passive WGs, which need to be protected. This leaves a SiO$_2$ layer covering as much area as possible, to ease transfer and preserve SLG continuity and integrity. 

Fig.\ref{fig3:SOI-Wafer}\textbf{b} shows a set of WGs prior to SLG transfer, where 1$\mu$m thick SiO$_2$ sleeves on the passive WGs can be seen. After SLG transfer, trenches are observed in the proximity of the sleeves, but SLG is continuous on the WG active area, Fig.\ref{fig3:SOI-Wafer}\textbf{c}, where the device is to be fabricated. Surface smoothness of the target substrate and residual SLG doping are key parameters to be controlled during device fabrication. Hence, we test 3 types of etching to understand which one produces the best SiO$_2$ surface in terms of roughness and residual SLG doping after transfer: reactive ion etching (RIE), hydrofluoric acid (HF), and RIE-inductively coupled plasma (RIE-ICP).

After CMP on a 200mm wafer to reduce the SiO$_2$ thickness to 1$\mu$m, we divide the wafer in 4 quarters, and etch 3, as mentioned above. Then, we collect atomic force microscopy (AFM, Bruker Icon) topography images and evaluate roughness over the WG profile for an area of 200$\times$200nm$^2$ on the Si core, Fig.\ref{fig3:SOI-Wafer}\textbf{d},\textbf{e}. We base our analysis on root mean square (RMS) roughness, defined as $R_{RMS}=\sqrt{\frac{\sum Z_i^2}{N}}$\cite{Whitehouse2004}, where $Z_i$ is the height of point $N$ within the evaluation area. We find comparable RMS roughness, with RIE etched surfaces slightly smoother than the others. with roughness 0.23$\pm$0.03nm. ICP-etched SiO$_2$ has the highest RMS roughness$\sim$0.31nm. This can be explained by ICP etching typically involving higher ion energies and flux densities than RIE\cite{Plummer2000}. 

We then transfer CVD SLG from$\sim$4" Cu foils on each quarter, and analyse the Raman peaks at 514.5nm for the RIE, HF and ICP cases (Fig.\ref{fig4r:Raman}\textbf{a-e}). We collect 16 spectra for each. Table\ref{tab2:RamanEtchTest} summarises the Raman peaks fits, E$_F$, doping type, $n$, strain, and defects density n$_D$. The derivation of material parameters from Raman spectra is described in Methods. Our experiments show a difference between SLG transferred on ICP-etched SiO$_2$ and RIE- or HF-etched SiO$_2$. SLG on ICP-etched SiO$_2$ has E$_F$=360$\pm$14 meV, $\sim$1.5 times higher than RIE-etched SiO$_2$, and twice HF-etched SiO$_2$. ICP has n$_D\sim$4.2 times higher than RIE and$\sim$5.3 times higher than HF. HF is slightly better than RIE in terms of E$_F$ (187$\pm$57, 234$\pm$99meV) and n$_D$ (0.68$\pm$0.21$\times$10$^{10}$, 0.85$\pm$0.70$\times$10$^{10}$cm$^{-2}$). However, the isotropic nature of HF and the high reactivity with both SiO$_2$ and Si\cite{Kang2002}, might lead to unwanted damage to the Si WGs. Due to the combination of low $n$ (13.7$\pm$13.2$\times$10$^{12}$cm$^{-2}$), low (0.85$\pm$0.70$\times$10$^{10}$cm$^{-2}$) n$_D$ and atomically smooth SiO$_2$ (see Fig.\ref{fig3:SOI-Wafer}\textbf{d} and parameters in Table\ref{tab2:RamanEtchTest}), RIE is the preferred etching method for SLG processing, and can be used to structure the SiO$_2$ topography before transfer on WGs, without compromising SLG/Si WGs properties.
\begin{figure*}
\centerline{\includegraphics[width=150mm]{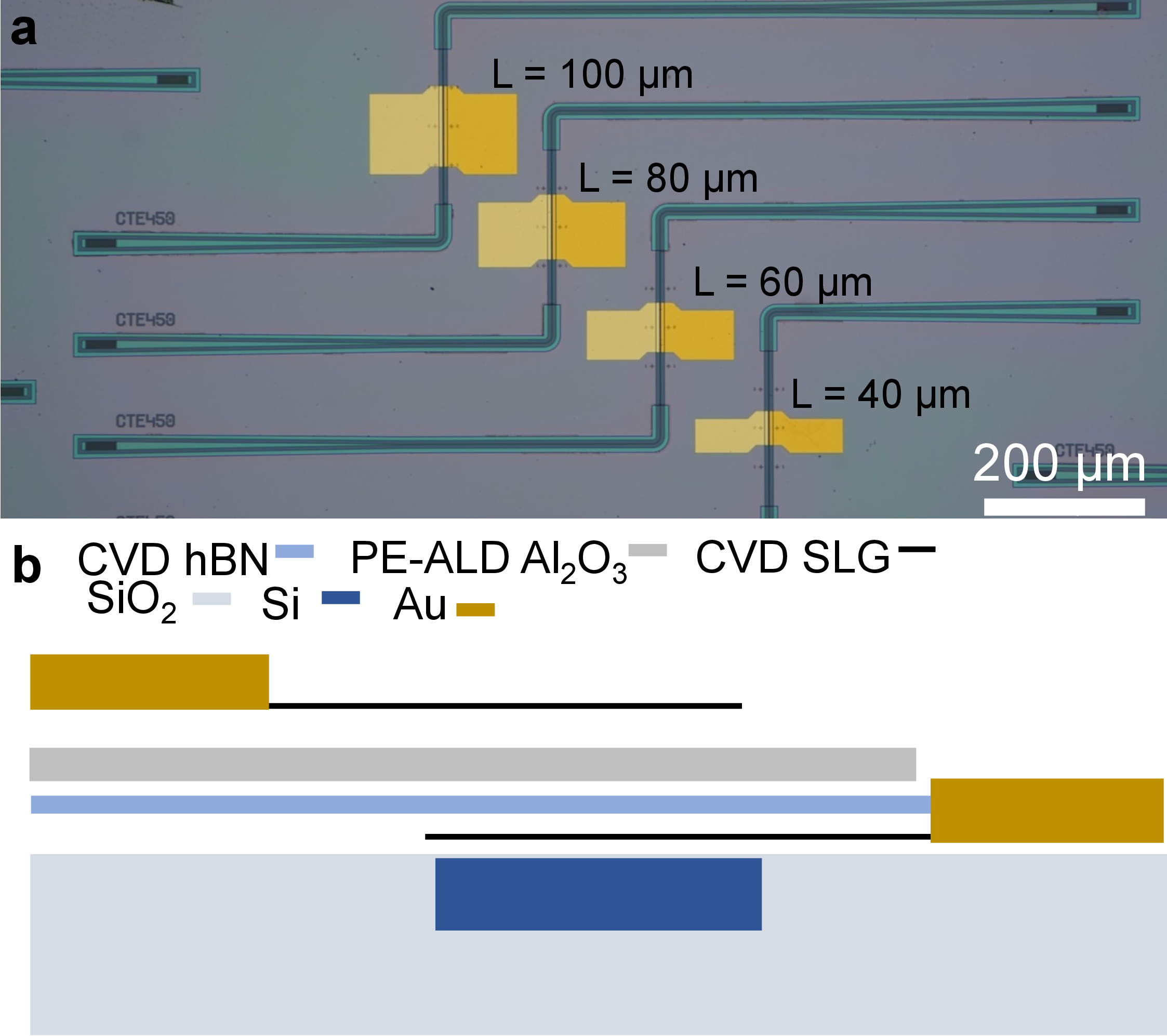}}
\caption{\textbf{a} Optical image of DSLG EAMs with different $L$. \textbf{b} Cross-section of DSLG EAMs with CVD hBN and Al$_2$O$_3$ compound gate dielectric.}
\label{fig4:DeviceImage}
\end{figure*}
\begin{figure*}
\centerline{\includegraphics[width=180mm]{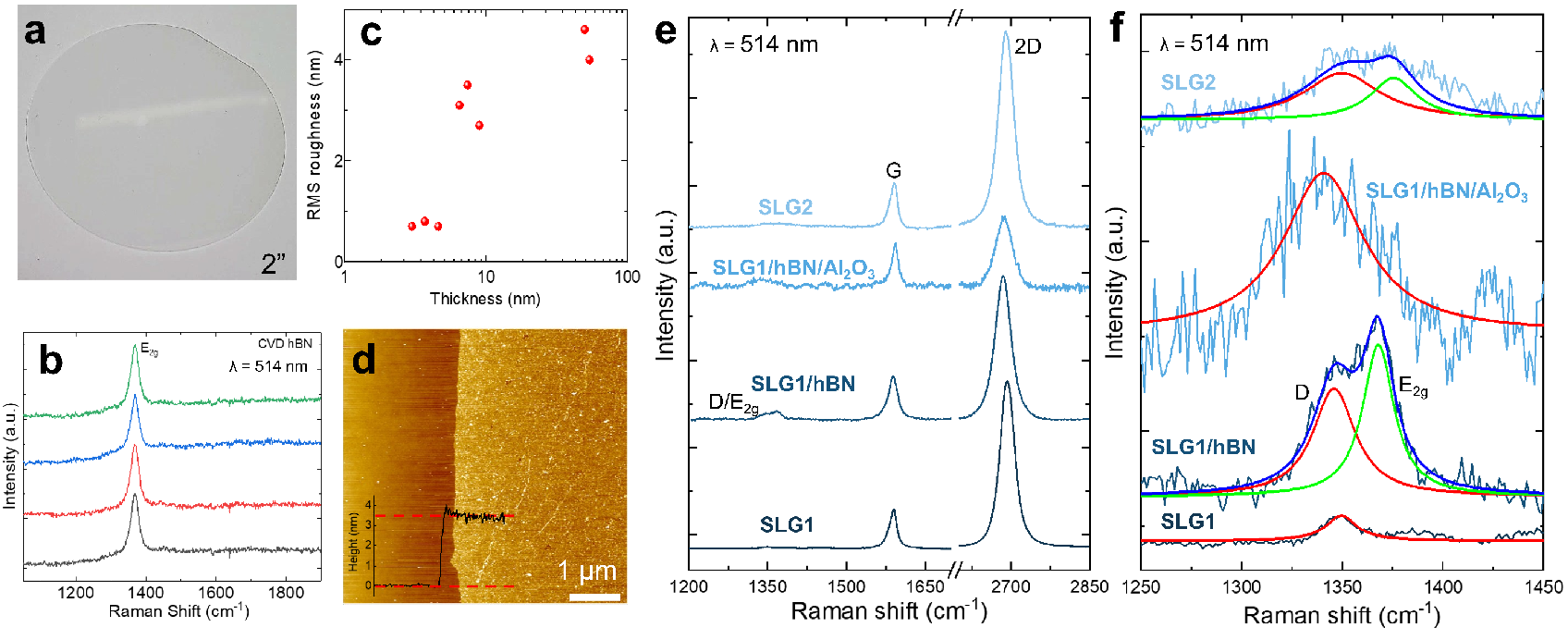}}
\caption{\textbf{a} 2" sapphire wafer with 3.5 nm thick hBN grown by MOCVD before transfer. \textbf{b} Raman spectra of hBN on Sapphire at 514 nm  \textbf{c} RMS roughness of CVD hBN as a function of thickness after transfer. \textbf{d} AFM image of the same hBN after transfer, showing a 3.5 nm step height.\textbf{e} Raman spectra of SLG1 and SLG2 between device fabrication steps. \textbf{f} Zoomed-in Raman spectra shown in \textbf{e}, indicating the hBN E$_{2g}$ peak ($\sim$ 1370 cm$^{-1}$) and the SLG D peak ($\sim$ 1349 cm$^{-1}$) at 514 nm. }
\label{fig5:MaterCar}
\end{figure*}
\begin{table*}[hbtp]
    \centering
    \begin{tabular*}{\linewidth}{@{\extracolsep{\fill}}c c c c c}
    \toprule 
     Samples & SLG1 on SOI & hBN on SLG1 & Al$_2$O$_3$ on hBN/SLG1 & SLG2 on Al$_2$O$_3$ \\ 
    \midrule 
    Pos(G) (cm$^{-1}$) & 1588 $\pm$ 2 & 1586 $\pm$ 1 & 1590 $\pm$ 2 & 1588 $\pm$ 2 \\
    FWHM(G) (cm$^{-1}$) & 17 $\pm$ 2 & 24 $\pm$ 6 & 27 $\pm$ 9 & 18 $\pm$ 3 \\
    Pos(2D) (cm$^{-1}$) & 2691 $\pm$ 4 & 2684 $\pm$ 1 & 2687 $\pm$ 2 & 2689 $\pm$ 2 \\ 
    FWHM(2D) (cm$^{-1}$) & 33 $\pm$ 2 & 34 $\pm$ 2 & 38 $\pm$ 2 & 35 $\pm$ 2 \\
    A(2D)/A(G) & 5.9 $\pm$ 0.9 & 3.3 $\pm$ 1.7 & 3.0 $\pm$ 0.7 & 3.8 $\pm$ 1.0 \\
    I(2D)/I(G) & 3.2 $\pm$ 0.7 & 3.8 $\pm$ 1.4 & 2.3 $\pm$ 0.3 & 3.7 $\pm$ 0.8 \\
    I(D)/I(G) & 0.05 $\pm$ 0.04 & 0.07 $\pm$ 0.04 & 0.10 $\pm$ 0.05 & 0.07 $\pm$ 0.07 \\
    E$_F$ (meV) & 126 $\pm$ 53 & 213 $\pm$ 133 & 293 $\pm$ 97 & 164 $\pm$ 76 \\
    Doping type & p & p & p & p \\
    n ($\times$ 10$^{12}$ cm$^{-2}$) & 1.9 $\pm$ 0.7 & 8.9 $\pm$ 7.6 & 12.7 $\pm$ 7.6 & 5.8 $\pm$ 4.0 \\
    Uniaxial strain (\%) & -0.18 $\pm$ 0.10 & -0.12 $\pm$ 0.04 & -0.30 $\pm$ 0.09 & -0.19 $\pm$ 0.09 \\
    Biaxial strain (\%) & -0.07 $\pm$ 0.04 & -0.05 $\pm$ 0.01 & -0.11 $\pm$ 0.04 & -0.07 $\pm$ 0.04 \\
    n$_D$ ($\times$ 10$^{10}$ cm$^{-2}$) & 1.3 $\pm$ 1.2 & 3.7 $\pm$ 2.6 & 5.1 $\pm$ 2.9 & 2.0 $\pm$ 2.0 \\
    \bottomrule 
    \end{tabular*}
    \caption{Raman peaks analysis and corresponding E$_F$, doping type, $n$, strain, n$_D$ and related uncertainties of SLG1 and SLG2 between fabrication steps.}
    \label{tab3:RamanDevFab}
\end{table*}

\section{Device fabrication}
Fig.\ref{fig4:DeviceImage}\textbf{a},\textbf{b} show a top-view of DSLG EAMs with different L, and a cross-section of the device discriminating between the materials employed in the fabrication process. SLG and hBN are transferred on passive WGs, with partially etched grating couplers for input and output coupling. Fully etched Si WGs are fabricated through multiple steps involving deep-UV projection lithography, followed by Si etching via ICP (Oxford Instruments ICP380), resist stripping, and cleaning\cite{Littlejohns2020}. A 1$\mu$m-thick SiO$_2$ top cladding is then deposited by plasma-enhanced CVD and planarized using CMP. To prepare the substrate for SLG transfer, the top cladding layer is selectively removed using a pattern defined by deep-UV lithography, and a combination of etching techniques. The platform consists of a 220nm thick Si waveguiding layer on a 2$\mu$m buried oxide layer.

After the first SLG transfer (SLG1, see Methods), electron beam lithography (EBPG Raith 5200) on a PMMA layer is used to shape the etch mask for SLG1, and O$_2$ plasma is then used to etch it. This defines the overlap of SLG1 with the WG and the length of the modulator. Metallisation is done by a two-step process. First,$\sim$10$\mu$m-wide contacts are defined by a double PMMA layer e-beam lithography process and 100nm Au is thermally evaporated (Moorfield Nanotechnology, MiniLab 60) to form Au/SLG junctions. Then, a second lithography step defines the larger pads used for probing, comprising 3nm Cr and 100nm Au, both thermally evaporated. The two-step metallisation process is crucial for achieving high EO-BW, because Au/SLG junctions have lower $R_C\sim$340$\Omega\mu$m\cite{Sundaram2011, Cusati2017} compared to other metals (e.g. Cr/Au, $R_C>$1k$\Omega\mu$m\cite{Nagashio2009, Song2012}), reducing the time constant of the modulator's circuit, see Fig.\ref{Fig2:BW}\textbf{a}.

We use multilayer ($\sim$10layers) hBN before plasma-enhanced atomic layer deposition (PE-ALD) of Al$_2$O$_3$ for protecting SLG from the plasma involved in the PE-ALD process\cite{Giambra2021, Canto2021}. We grow hBN by metal-organic chemical vapour deposition (MOCVD, Aixtron CCS2D) on a 2" Sapphire wafer (Fig.\ref{fig5:MaterCar}\textbf{a})\cite{Calandrini2023}. After cleaving hBN on Sapphire into smaller dices ($<$1cm$^2$), we delaminate hBN as for Methods. Fig.\ref{fig5:MaterCar}\textbf{b} shows the hBN Raman E$_{2g}$ peak$\sim$1368cm$^{-1}$\cite{Reich2005} at 514nm. We investigate the roughness of hBN after transfer and find a thickness dependence, Fig.\ref{fig5:MaterCar}\textbf{c}. For hBN$<$6nm, we have$<$1nm roughness, while$>$10nm-thick we get$>$2nm roughness. Hence, to maintain low ($<$1nm) roughness throughout the process, we use a 3.5nm-thick hBN (Fig.\ref{fig5:MaterCar}\textbf{d}), which guarantees both protection from plasma and low roughness.

We deposit 40nm Al$_2$O$_3$ on hBN by PE-ALD (Fiji, Veeco) at T=150°C using tri-methyl-aluminium (TMA) and oxygen precursors. Unlike SiO$_2$ and Si$_3$N$_4$, which require 400°C for PECVD\cite{Plummer2000}, Al$_2$O$_3$ can be deposited at lower T=150°C by PE-ALD\cite{Plummer2000}. Reducing the processing T minimizes the likelihood of damaging SLG/hBN and improves compatibility with CMOS backend-of-line (BEOL) processing, which has a thermal budget of 450°C\cite{Franke1999}. We then repeat SLG transfer, shaping, etching and metallisation to complete the DSLG structure.

For driving DSLG EAMs at 1.55$\mu$m, a doping corresponding to E$_F=$400meV (see Fig.\ref{Fig1:EvanescentCoupling}\textbf{c}) is beneficial,  because no DC bias would be required for driving the EAMs. However, a residual E$_F\sim$400meV is typically related to charged impurities\cite{Chen2008} and defects\cite{Bruna2014}, reducing $\mu$\cite{Chen2008, Ni2010}, increasing IL, and decreasing EO-BW. Hence, we perform Raman spectroscopy after each fabrication step to monitor how this affects SLG1 and SLG2, Fig.\ref{fig5:MaterCar}\textbf{e}. Table \ref{tab3:RamanDevFab} summarises Raman peaks fits, E$_F$, doping type, $n$, strain, and n$_D$ after each step. SLG1 has E$_F$=126$\pm$53meV after transfer on SOI, increasing to 213$\pm$133meV after hBN encapsulation, and to 293$\pm$97meV after Al$_2$O$_3$ deposition. The increase in E$_F$ occurs with an increase in n$_D$ and an increase in strain.

SLG1 degradation, due to contaminations and stress related to consecutive fabrication steps, such as hBN wet transfer and PE-ALD deposition of Al$_2$O$_3$, is visible from the spectra in Fig.\ref{fig5:MaterCar}\textbf{e}, which show broadening of G and 2D peaks, reduction in I(2D)/I(G), and increase in I(D)/I(G). The double Lorentzian peak$\sim$1360cm$^{-1}$ in Fig.\ref{fig5:MaterCar}\textbf{e}, \textbf{f} for the SLG1/hBN spectrum is due to the overlap of the SLG1 D peak and hBN E$_{2g}$ peak at 514nm\cite{Purdie2018}. Fig.\ref{fig5:MaterCar}\textbf{f} shows how the SLG1 D and hBN E$_{2g}$ peaks can be identified and fitted using two Lorentzians, except for the SLG1/hBN/Al$_2$O$_3$ spectrum where the D peak dominates over E$_{2g}$. The final E$_F$=293$\pm$97meV and n$_D$=5.1$\pm$2.9$\times$10$^{10}$cm$^{-2}$, despite increasing due to the fabrication process, are suitable for operating the device as EAM, because E$_F$ is close enough to the quadrature point, E$_F\sim$400meV, to induce Pauli blocking, maximizing ER/V and minimizing IL, while the low n$_D$ minimizes electron scattering due to defects\cite{Ni2010}. SLG2 does not undergo the same fabrication steps, and has lower E$_F$=164$\pm$76meV, n$_D$ and strain (see Table\ref{tab3:RamanDevFab}).

\begin{figure*}
\centerline{\includegraphics[width=180mm]{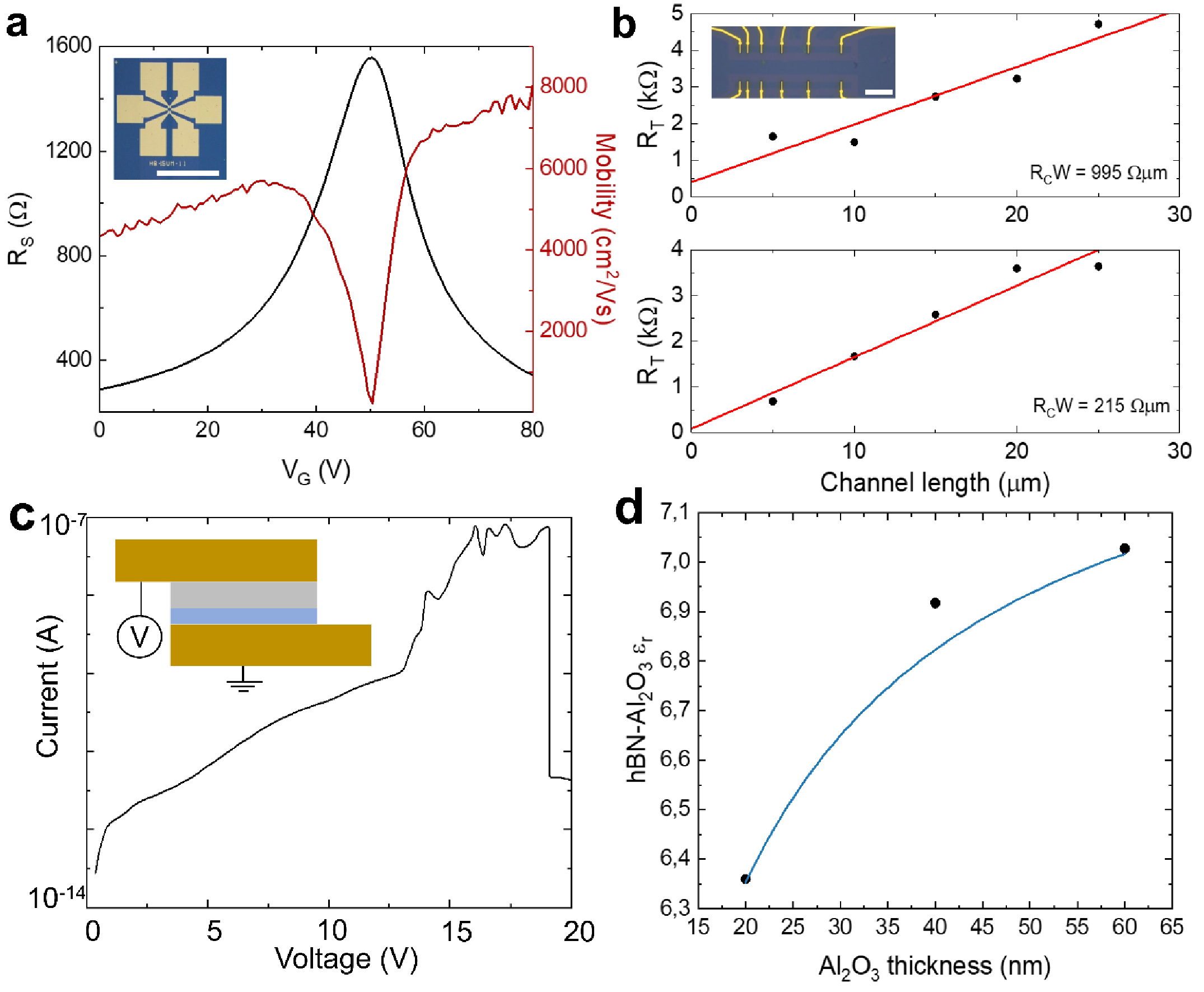}}
\caption{\textbf{a} R$_S$ and $\mu$ as a function of back-gate voltage, extracted from the GFETs shown in the inset. Scale bar: 500$\mu$m. \textbf{b} Total resistance as a function of channel length of SLG TLM structures fabricated with Au/SLG junctions, shown in the inset. Scale bar: 10$\mu$m. The contact resistance extrapolated is between 215 and 995$\Omega\mu$m. \textbf{c} I-V curve of a 600$\mu$m$^2$ hBN/Al$_2$O$_3$ capacitor between Au electrodes showing dielectric breakdown at 19V. \textbf{d} Extrapolation of relative permittivity from capacitance measurements of hBN-Al$_2$O$_3$ capacitors with different Al$_2$O$_3$ thicknesses.}
\label{fig6:ElCal}
\end{figure*}

Fig.\ref{fig6:ElCal}\textbf{a}, \textbf{b} plots 4-point-probe measurements of a back-gated Hall bar, and the total resistance as a function of channel length in transfer length method (TLM\cite{Schroder2006}) structures, made to characterise $\mu$ and contact resistance of the SLG used to fabricate the modulators. We get $\mu\sim$8,000cm$^2$/Vs. TLM measurements are done on Au/SLG junctions, revealing a maximum contact resistance$\sim$995$\Omega$$\mu$m and a minimum$\sim$215$\Omega$$\mu$m. The extraction of $\mu$ and $R_C$ from electrical test structures is described in Methods. 

We characterize the Al$_2$O$_3$/hBN gate dielectric in terms of breakdown electric field and relative permittivity as a function of Al$_2$O$_3$ film thickness, Fig.\ref{fig6:ElCal}\textbf{c},\textbf{d}. The relative permittivity is first extracted by measuring capacitance as a function of area, and then estimated for different Al$_2$O$_3$ thicknesses as described in Methods. We get $\epsilon_r\sim$6.35 for a thickness of 20nm, $\sim$6.9 for 40nm, and 7.05 for 60nm (see Fig.\ref{fig6:ElCal}\textbf{d}). We measure up to 0.95Vnm$^{-1}$ breakdown electric field, corresponding to$\sim$20V for 20nm thick dielectric and$\sim$40V for 40nm. The measured breakdown electric field (0.95Vnm$^{-1}$) and relative permittivity ($\sim$6.9 for 40nm Al$_2$O$_3$) are similar to the state-of-the-art for other PE-ALD Al$_2$O$_3$ films\cite{Illarionov2020, Lim2004}, confirming the suitability of Al$_2$O$_3$ as a CMOS BEOL-compatible dielectric. This enables us to probe the transparent regime\cite{Watson2023} and reduce IL\cite{Sorianello2018, Watson2023}.
\begin{figure*}
\centerline{\includegraphics[width=180mm]{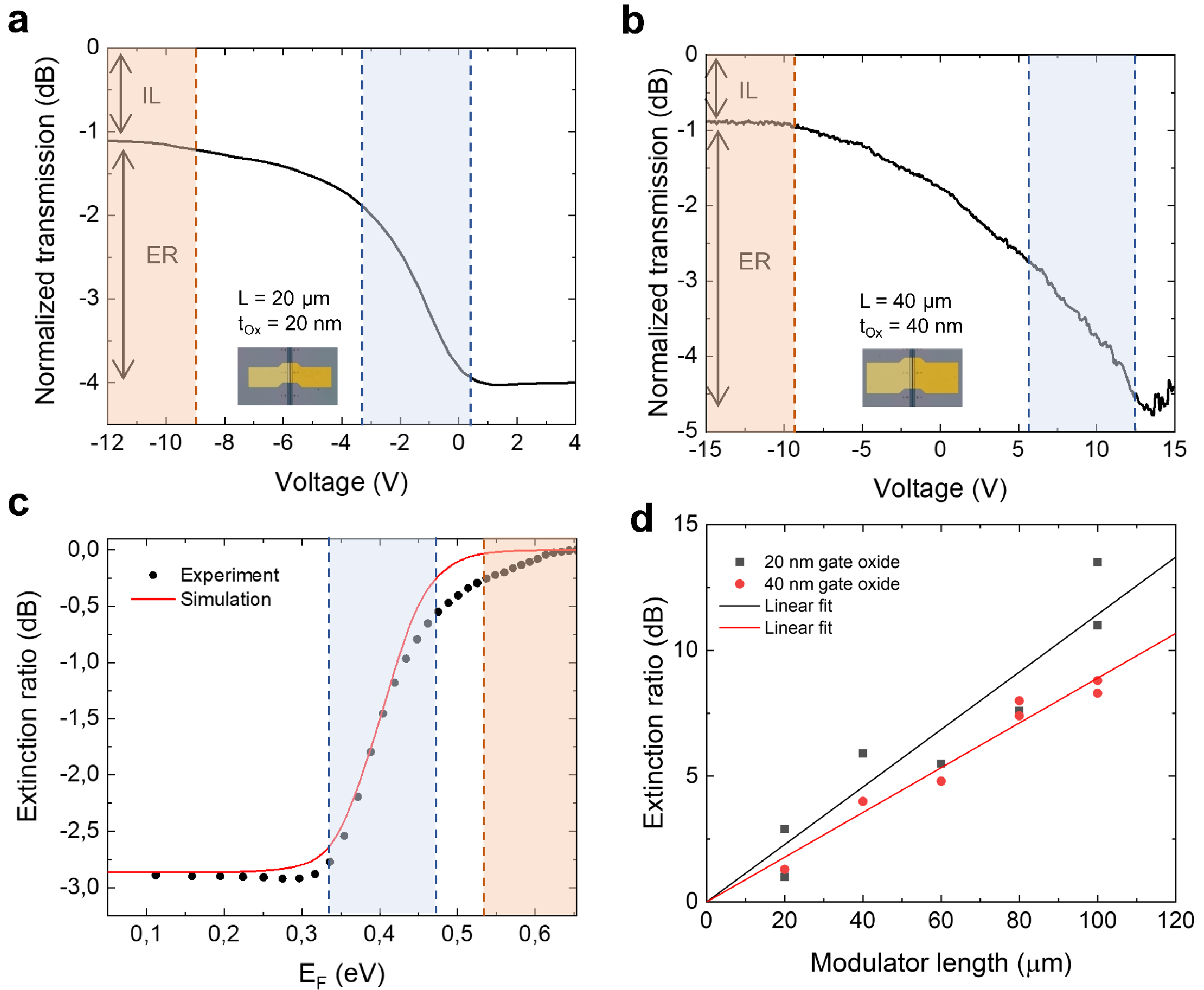}}
\caption{\textbf{a} Optical transmission normalized to device optical loss before fabrication as a function of DC bias voltage in the C-band (1550nm) for \textbf{a} L=20$\mu$m and 20nm gate dielectric design, and \textbf{b} L=40$\mu$m with 40nm gate dielectric. The blue region indicates the working point for electro-absorption modulation, which is$\sim$0.4eV for $\lambda$=1550nm. The orange region is the transparent regime, where interband transitions do not occur due to Pauli blocking and only intraband transitions contribute to absorption. \textbf{c} Simulated and experimental ER as a function of E$_F$ for L=20$\mu$m. A maximum E$_F$=0.64eV is achieved. The orange region illustrates the transparency regime (Pauli blocking). \textbf{d} Measured static ER for modulators with different lengths. The slope is 0.12dB/$\mu$m for the 20nm gate dielectric design, while it is 0.09dB/$\mu$m for the 40nm one}
\label{fig5:EO-response1}
\end{figure*}
\begin{figure*}
\centerline{\includegraphics[width=180mm]{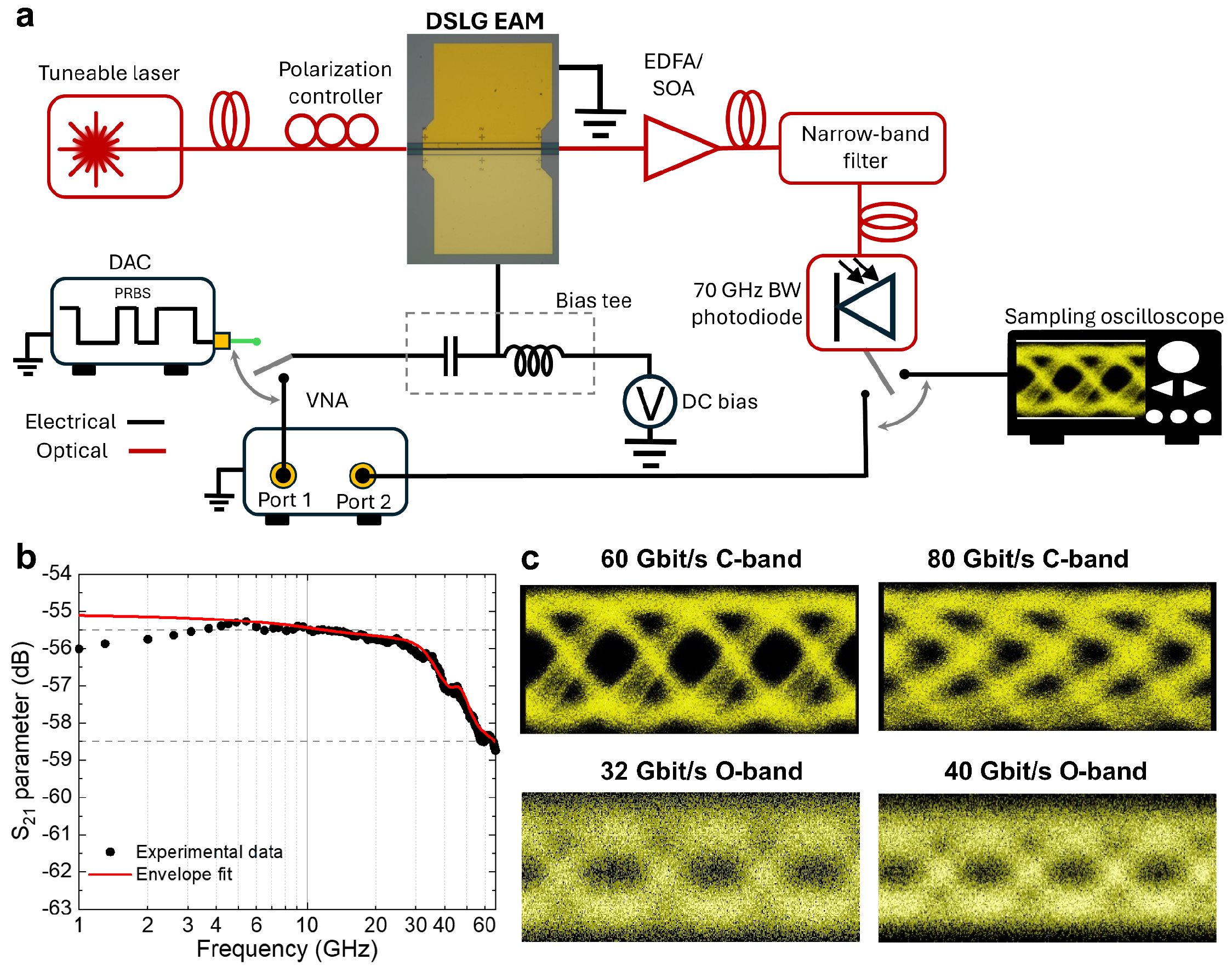}}
\caption{\textbf{a} Set-up for dynamic characterization showing the DSLG EAM under test. \textbf{b} S$_{21}$ parameter as a function of frequency showing a 3dB cut-off EO-BW$=$67GHz for a 40$\mu$m-long modulator (40nm gate oxide). \textbf{c} Eye diagrams of the same 40$\mu$m-long modulator showing 60 and 80Gbit/s NRZ data rate in the C-band, and a different 40$\mu$m-long modulator on the same chip showing 32 and 40Gbit/s NRZ data rate in the O-band. A SOA is used instead of an EDFA to amplify the signal in the O-band.}
\label{fig6:EO-response2}
\end{figure*}
\begin{table*}[bt]
    \centering
    \begin{tabular*}{\linewidth}{@{\extracolsep{\fill}}c c c c c c c c c c}
    \toprule 
     Length[$\mu$m] & NRZ rate[Gbit/s] & \multicolumn{2}{c}{SNR} & \multicolumn{2}{c}{BER} & \multicolumn{2}{c}{ER[dB]} \\ 
    \midrule 
    & & Original & Filtered & Original & Filtered & Original & Filtered \\
    \midrule
    40 & 50 & 3.25 & 5.17 & 5.7$\times$10$^{-4}$ & 1.2$\times$10$^{-7}$ & 1.20 & 1.05 \\
    40 & 60 & 2.63 & 4.02 & 4.1$\times$10$^{-3}$ & 2.9$\times$10$^{-5}$ & 1.19 & 1.02 \\
    40 & 80 & 1.91 & 2.38 & 2.8$\times$10$^{-2}$ & 8.5$\times$10$^{-3}$ & 1.16 & 0.82 \\
    80 & 80 & 1.66 & 2.06 & 4.8$\times$10$^{-2}$ & 1.9$\times$10$^{-2}$ & 1.22 & 1.17 \\
    100 & 50 & 2.95 & 4.84 & 1.6$\times$10$^{-3}$ & 6.6$\times$10$^{-7}$ & 1.38 & 1.89 \\
    100 & 60 & 2.41 & 3.75 & 7.9$\times$10$^{-3}$ & 8.9$\times$10$^{-5}$ & 1.33 & 1.74 \\
    100 & 80 & 1.76 & 2.31 & 3.8$\times$10$^{-2}$ & 1$\times$10$^{-2}$ & 1.27 & 1.41 \\
    \midrule
    Back-to-back & 60 & 3.76 & 5.32 & 8.5$\times$10$^{-5}$ & 5.1$\times$10$^{-8}$ & - & - \\
    Back-to-back & 80 & 2.31 & 2.70 & 1$\times$10$^{-2}$ & 3.4$\times$10$^{-3}$ & - & - \\
    \bottomrule 
    \end{tabular*}
    \caption{Summary of data transmission results extracted from DSLG EAMs eye diagrams with different L, with and without CTLE filtering. DAC back-to-back signal at 60 and 80Gbit/s is also shown.}
    \label{tab4:DataTransmissionTable}
\end{table*}

\section{Electro-optic characterization}
The electro-optic response of the DSLG EAMs is then characterized in both static (DC) and dynamic (DC+RF) operation. We use grating couplers for both input and output optical coupling to the WGs hosting the EAMs, with insertion loss$\sim$6dB per coupler, extracted from straight reference WGs without active devices. Coupling is performed using angled single-mode fibers on a probe station. The positions of fibres and polarization are adjusted to reduce coupling losses. We first use a tuneable laser (Agilent 8164B Lightwave Measurement System) with $P_{in}$=1mW and sweep the input wavelength in the C-band. We then measure the power at the output of the DLSG EAM, and find a maximum for $\lambda$=1.55$\mu$m. We then apply a DC voltage to the modulator.

Fig.\ref{fig5:EO-response1} shows the optical transmission of (\textbf{a}) 20$\mu$m long modulator with 20nm thick dielectric as a function of voltage, and (\textbf{b}) 40$\mu$m modulator with 40nm dielectric as a function of voltage, (\textbf{c}) change in ER as a function of E$_F$ of the combined SLGs for \textbf{a}, and (\textbf{d}) ER as a function of L for 20 and 40nm gate oxides. We measure devices with L=20, 40, 60, 80, 100$\mu$m and estimate ER$\sim$0.12dB/$\mu$m for 20nm oxide and$\sim$0.09dB/$\mu$m for 40nm (see Fig.\ref{fig5:EO-response1}\textbf{d}), corresponding to maximum SLG absorption with cladding thickness t$_{cladding}$=10nm (Fig.\ref{Fig1:EvanescentCoupling}\textbf{d}). 

This confirms that planarization leads to a$\sim$10nm thick cladding. For the 20nm dielectric device, output power can be modulated from $\sim$-1.1 to -4.1dB, corresponding to a maximum ER$\sim$3dB within a voltage sweep$\sim$10V. For the 40nm dielectric device, the output power can be modulated from$\sim$-0.1 to -4.6dB, corresponding to ER$\sim$4.5 dB within a voltage sweep$\sim$20V. 

The highlighted blue region in Fig.\ref{fig5:EO-response1}\textbf{a},\textbf{b} shows the steepest change in optical transmission due to Pauli blocking, centred at the quadrature point E$_F=$400meV, while the orange region corresponds to the transparency regime, where there is no AM. We get a modulation efficiency$\sim$0.037dB/V$\mu$m between 0.1 and -1.8V, with ER$\sim$1.4dB, after normalizing to L=20$\mu$m. This is the highest absorption modulation reported to date, to the best of our knowledge, for scalable SLG on SiPh.

For the 40nm thick dielectric device, the modulation efficiency goes down to$\sim$0.01dB/V$\mu$m between 10.5 and 13V, where the device exhibits ER$\sim$1dB, due to SLG2 being further away from the Si WG, and reduced gating efficiency. We extract E$_F$ from the gate-induced charge carriers density $n=(C_{ox}/e)(V_G-V_{Dirac})$, corresponding to a shift in $E_F=\hbar v_F\sqrt{n\pi}$\cite{Novoselov2004}. From static optical transmission modulation measurements, we get E$_F$=0.4eV at V$_G$=-1.8V, i.e. the point at which Pauli blocking begins and ER is halved. Then, we calculate $n=(E_{F}^{2})/(\hbar^2 v_{F}^{2}\pi)\sim1.2\times10^{13}cm^{-2}$ for E$_F$=0.4eV. We use this to derive $V_{Dirac}=V_{G}-en/C_{ox}\sim-4.2$V, with $C_{ox}=0.0031$Fm$^{-2}$ the geometric capacitance. Once the Dirac point is fixed, we have E$_F$ as a function of $n$, hence we can plot ER as a function of E$_F$, Fig.\ref{fig5:EO-response1}\textbf{c}. We measure E$_F$=0.64eV, $\sim$100meV higher than in Ref.\cite{Agarwal2021}, enabling access to the transparency region where static ER$\sim$0dB, thanks to the large breakdown electric field for our hBN-Al$_2$O$_3$ capacitor: $E_{break}=$0.95V/nm for 20nm thickness. The maximum E$_F$ can be calculated from $E_F^{Max}=\hbar v_F\sqrt{(\pi\epsilon_0\epsilon_r E_{break}/e)}$\cite{Agarwal2021}. We get E$_F^{Max}$=0.71eV.

The combination of SLG1 and SLG2 residual doping extracted from the modulator measurement is E$_F\sim$320meV, within the range extracted by Raman in Fig.\ref{fig5:MaterCar}\textbf{e},\textbf{f}, i.e. 230$\pm$120meV.
\section{Bandwidth and data transmission}
We then perform RF measurements using the setup in Fig.\ref{fig6:EO-response2}\textbf{a}. Similar to the DC characterization, a CW laser source is coupled to the device via vertical coupling with the GC input. The optical output of the modulator is then sent to an Er doped fiber amplifier (EDFA) to compensate for the GC losses and for those introduced by the fibres in the set-up ($\sim$3dB). The EDFA output is then filtered using a$\sim$1nm narrow-band filter to remove the out-of-band noise generated by the amplified spontaneous emission of the optical amplifier. The optical signal is then coupled to a 70GHz BW photodiode (Finisar XPD3120R). The DSLG modulator is electrically contacted with a 67GHz RF electrical probe in groud-signal (GS) configuration (FormFactor GS Infinity Porbe). A bias tee with BW$\sim$67GHz is connected to the probe to allow simultaneous DC and AC coupling. The DC bias is used to set the modulator working point, while the AC port is connected to either a VNA (Keysight PNA N5227B) for frequency response measurements, or to a Digital-to-Analog Converter (Micram DAC4) for eye diagram measurements, see Methods.

We start by measuring the EO frequency response of the device. After VNA calibration, with reference to Fig.\ref{fig6:EO-response2}\textbf{a}, port 1 of the VNA is connected to the DSLG modulator, while port 2 to the photodiode. The measured S$_{21}$ shows the modulator EO-BW. Fig.\ref{fig6:EO-response2}\textbf{b} plots S$_{21}$ for a 40 nm thick dielectric modulator with L=40$\mu$m, showing $f_{3dB}=$67GHz. The data are fitted with an upper envelope function after calibration with respect to the photodiode and probes frequency response. $f_{3dB}$ is extrapolated from -55.5 to -58.5dB, as indicated by the dashed grey lines in Fig.\ref{fig6:EO-response2}\textbf{b}. This is the highest $f_{3dB}$ reported for DSLG modulators to date, to the best of our knowledge. Our $f_{3dB}$ is 2.3 times higher than previous scalable DSLG EAMs\cite{Giambra2019} and 1.7 times higher than DSLG based on exfoliated flakes\cite{Agarwal2021}. 

The measurements in Fig.\ref{fig6:EO-response2}\textbf{b} validate the electrical circuit model of Fig.\ref{Fig2:BW}\textbf{e}, which predicts $f_{3dB}\sim$70GHz for L=40$\mu$m. Our $f_{3dB}$ is on par with Ge EAMs\cite{Liu2022, Hu2022}. However, our devices are not limited to the C- or L- bands, but can work in the O-band as well, making them useful for data centres interconnections and access networks. Compared with III-V EAMs\cite{Shahin2019}, our DSLG EAMs have twice BW. Our DSLG EAM $f_{3dB}$ lags behind Si MZM\cite{Han2023} and LN MZMs\cite{He2019}, but offers the advantage of smaller footprint (22$\mu$m$^2$ against$\sim$104$\mu$m$^2$\cite{Han2023} and 5000$\mu$m$^2$\cite{He2019}) and lower energy consumption per bit ($\sim$58fJ/bit against$\sim$170fJ/bit\cite{He2019}). 

We then perform NRZ eye diagram measurements by connecting the DAC to the DSLG modulator, as in Fig.\ref{fig6:EO-response2}\textbf{a} and described in Methods. Given the high speed of the device under test (DUT), revealed by frequency response characterization, this measurement targets the highest data rate reachable by our setup (80Gbit/s NRZ). It is therefore crucial to compensate for all the in-band losses and distortions due to non-flat frequency response of the measurement chain. This comprises DAC, electrical amplifier, one RF cable, bias tee, GS probe, photodiode, and oscilloscope. The compensation is performed by pre-emphasizing the DAC sequence, i.e. shaping the signal to boost the DAC high-frequency amplitude\cite{Buckwalter2006}, and after sequence acquisition with the oscilloscope. Feed-forward equalizers (FFE) filters and CTLE filtering\cite{Francese2015} are implemented, allowing lower frequency components ($<$10GHz) attenuation and higher frequency components enhancement\cite{Francese2015} (around the Nyquist frequency, defined as one-half of the sampling rate, $f_{\text{N}}=f_{\text{s}}/2$\cite{Oppenheim1999}). The overall effect is a flattening of the frequency response of the measurement chain inside the band of interest, at the expense of peak-to-peak voltage reduction of the electrical signal applied to the DUT, and of the acquired signal. Fig.\ref{fig6:EO-response2}\textbf{c} shows data rates of 60 and 80Gbit/s. To the best of our knowledge, these are the highest reported to date for DSLG modulators, surpassing Ref.\cite{Giambra2019} by 30Gbit/s and Ref.\cite{Agarwal2021} by 40Gbit/s. Substituting the C-band laser and EDFA with a O-band laser and a semiconductor optical amplifier (SOA), we measure up to 40Gbit/s in the O-band, demonstrating the SLG potential for high-speed broadband operation. The devices in the O-band are fabricated with the same methodology on the same chip, with Si WGs width 400 instead of 450nm, to optimize O-band transmission.

Table \ref{tab4:DataTransmissionTable} shows NRZ data rates of several modulators working in the C-band, together with their SNR, BER, and ER, as well as the electrical signal generated by the DAC and fed into the modulators, before and after CTLE filtering. SNR, BER, ER decrease as data rate increases, due to increasing noise in the data generation from the DAC (MicramDAC4, with sampling rate 100Gs). The SNR of the DAC output after filtering is halved when going from 60 to 80Gbit/s, see Table\ref{tab4:DataTransmissionTable}. This explains why this eye degradation also occurs for the modulator, with BW in principle allowing for larger data rates to be transmitted, resulting in smaller ER and BER. For modulators with L=40$\mu$m, we get BER$\sim$1.2$\times$10$^{-7}$ at 50Gbit/s, decreasing to$\sim$8.5$\times$10$^{-3}$ at 80Gbit/s. The reason for this BER is two-fold. First, the original electrical BER$=1\times10^{-2}$ at 80Gbit/s would increase with a measurement setup providing higher SNR. Second, the DSLG EAM with 40nm gate dielectric, while providing a broad EO-BW$=$67GHz, has poor modulation efficiency$\sim$0.01dB/V$\mu$m, hence we apply V$_{pp}\sim$7V to drive the modulator, achieving ER$\sim$1.2dB. Nonetheless, our work demonstrates the fastest DSLG EAMs to date, and a fabrication flow to achieve designs with high (67GHz) BW and static modulation efficiency in DSLG EAMs. A better control over SLG doping after transfer, would reduce sheet and contact resistances, without affecting $\mu$. E.g., for a DSLG EAM with a 20nm dielectric, R$_{ungated}\sim$27$\Omega/\square$ (corresponding to E$_F\sim$0.6eV), R$_C\sim$200$\Omega\mu$m. For a 450nm-wide gated SLG region with R$_{gated}\sim$60$\Omega/\square$, a EO-BW f$_{3dB}\sim$90GHz is possible, with $\mu\sim$8,000cm$^2$/Vs, without compromising modulation efficiency, making graphene integrated amplitude modulators an attractive technology for high-speed optical communications and computing, particularly for$>$1Tb/s IM-DD with DSP-free NRZ transceivers.
\section{Conclusions}
We reported DSLG EAMs prepared with a scalable fabrication process. The modulators are fabricated on a dedicated SOI wafer with optimised planarization to ease SLG and hBN integration, resulting in improved device performance. We used a hBN/Al$_2$O$_3$ gate dielectric with high breakdown electric field$\sim$0.95V/nm, enabling E$_F$ tuning of 0.64eV by solid-state gating. The access to SLG transparency, together with the enhanced SLG-mode interactions due to the planarization, enables average ER$\sim$0.12dB/$\mu$m and modulation efficiency$\sim$0.037dB/V$\mu$m for modulators with 20nm gate dielectric. The selective planarization produced a structured cladding where the passive regions do not interact with SLG and fabrication-related contaminants, hence reducing IL. We demonstrated the fastest graphene amplitude modulators to date, with EO-BW f$_{3dB}=$67GHz and 80Gbit/s NRZ transmission, with low energy consumption ($\sim$58fJ/bit), low insertion loss (IL$\sim$0.9dB), and small footprint ($\sim$22$\mu$m$^2$). Our work paves the way to foundry-ran graphene integrated photonics MPW for a variety of applications in optical communications, computing, and sensing, which require low-loss ($\sim$0.05dB/$\mu$m), ultra-fast (67GHz), and broadband (operational in both O-band and C-band) optical modulators. Our DSLG EAMs can be used to explore Tbit/s DSP-free NRZ transceivers for data center interconnects, reducing system-level energy consumption, beneficial for processing AI workloads.
\section{Methods}
\textbf{Material preparation}. SLG is grown on a polycrystalline Cu foil (25$\mu$m, Graphene Platform, Japan) by hot wall chemical vapour deposition (CVD)\cite{Li2009cvd}. The Cu foil is first annealed at 1050${^o}$C for 2h at atmospheric pressure (50sccm H$_2$ and 500sccm Ar, 760Torr), then annealed in pure H$_2$ (40sccm,$\sim$0.4Torr) for 3h using a contact-free method\cite{Jin2018}. Next, 5sccm CH$_4$ and 40sccm H$_2$ are introduced for 30mins to grow the SLG film, followed by turning off all gases and the heater to cool the sample ($\sim$1mTorr) to RT in vacuum. SLG is then separated from the Cu foil by electrochemical delamination using PMMA as supporting layer\cite{Chen2017}, and left floating in ultrapure DI water. Then, the target substrate is brought into contact with the SLG/PMMA stack in water and dried in air. After drying and removal of the supporting PMMA, SLG covers the whole target surface.

3.5nm hBN is grown on sapphire in an Aixtron CCS 2D reactor at 1400°C and 500mbar using borazine as precursor, carried by 10sccm N$_2$\cite{Calandrini2023}. Before growth, the sapphire is annealed in H$_2$ at 1200°C and 150mbar for 10mins\cite{Calandrini2023}. The wet delamination of hBN from sapphire is done as follows. A PMMA A4 950 layer is spin coated at 1000rpm on sapphire/hBN and baked for 10mins to serve as supporting layer. Then, we place sapphire/hBN/PMMA in a diluted orthophosphoric acid solution and heat to 50°C to facilitate the intercalation process. Once hBN/PMMA is fully detached from sapphire, it is transferred in a ultrapure DI water bath for cleaning, and then transferred on the target substrate.

\textbf{Raman characterization} Raman measurements are performed with a InVia spectrometer equipped with a 50x objective with NA=0.75 and 1cm$^{-1}$ resolution. Errors are estimated from the standard deviation across different spectra, the spectrometer resolution and the uncertainty associated with the different methods to estimate the doping from full width at half maximum of G-peak, FWHM(G), intensity and area ratios of 2D and G peaks, I(2D)/I(G), A(2D)/A(G)\cite{Pisana2007, Das2008, Basko2009}. $n$ is determined from A(2D)/A(G), I(2D)/I(G), and FWHM(G) as for Refs.\cite{Das2008, Basko2009}. Pos(2D)'s doping-dependent behavior determines whether SLG is p-doped or n-doped\cite{Das2008}. Strain is derived from Pos(G), as follows. First, $E_F$ is derived from A(2D)/A(G), I(2D)/I(G), and FWHM(G)\cite{Pisana2007, Das2008, Basko2009}. Then, the Pos(G) corresponding to this $E_F$ is calculated\cite{Mohiuddin2009}. Strain is determined from the difference between experimental and calculated Pos(G): (Pos(G)$_{calc}$-Pos(G)$_{exp}$)/($\Delta$Pos(G)),with $\Delta$Pos(G)$\sim$23cm$^{-1}$/\% for uniaxial strain and$\sim$60cm$^{-1}$/\% for biaxial\cite{Mohiuddin2009}. $n_D$ is derived from I(D)/I(G) for a specific $E_F$, using n$_D$=(2.7$\pm$0.8)$\times$10$^{10}$E$_L^4$[eV]I(D)/I(G){E$_F$[eV]}$^{(0.54\pm 0.04)}$\cite{Bruna2014}. In all spectra, the 2D peak is a single-Lorentzian, signature of SLG\cite{Ferrari2006,Ferrari2013}. No D peak is observed in SLG on Cu, indicating negligible Raman active defects\cite{Ferrari2000, Cancado2011}.

\textbf{SLG mobility, sheet resistance, and contact resistance extraction.} $\mu$ is extracted using four-probe measurements. This allows the intrinsic $R_S$ of SLG to be isolated, critical for accurate $\mu$ extraction\cite{Nagashio2011}. The conductivity $\sigma$ is then related to $n$ via the Drude model\cite{Drude1900}:
\begin{equation}
    \mu = \frac{\sigma}{n e} = \frac{1}{R_s n e} ,
\end{equation}
while \textit{n}, induced by $V_G$, is determined by $C_{ox}=\frac{\epsilon_0\epsilon_r}{d_{ox}}$\cite{Das2008}:
\begin{equation}
    n = \frac{C_{ox}}{e}(V_G - V_{Dirac}) ,
\end{equation}
where the Dirac voltage, $V_{Dirac}$, is the charge neutrality point. The field effect $\mu$ is then extracted from Eqs.4,5:
\begin{equation}
    \mu=\frac{1}{C_{ox}}\frac{d\sigma}{dV_G}.
\end{equation}

TLM structures are employed to quantify $R_C$ between SLG and metal contacts. TLM is based on measuring the total resistance of SLG channels of varying lengths. The total resistance is modeled as\cite{Sze2006}:
\begin{equation}
    R_{tot} = 2R_C +\frac{L}{W}R_S,
\end{equation}
with $L$ the channel length and $W$ the width. By plotting $R_{tot}$ as a function of $L$, the slope of the linear fit gives $R_S/W$, while the intercept at $L= 0$ gives 2$R_C$.

\textbf{Gate dielectric characterization}. hBN-Al$_2$O$_3$ test capacitors between Au electrodes of different size (from 600 to 3600$\mu$m$^2$) and thickness (20, 40, 60nm) are fabricated using the same transfer, patterning, and deposition methods. We measure the breakdown electric field by applying a bias to signal and ground pads of the capacitor, and we extract the relative permittivity via one-port (S$_{11}$) measurements with frequency$<$1GHz, Fig.\ref{fig6:ElCal}\textbf{c},\textbf{d}. The relative permittivity is first extracted by measuring capacitance as a function of area, then estimated for different Al$_2$O$_3$ thicknesses. We fit the data with $\epsilon_r$=$\frac{(d_{hBN}+d_{Al_2O_3})\epsilon_{hBN}\epsilon_{Al_2O_3}}{\epsilon_{hBN}d_{Al_2O_3}+\epsilon_{Al_2O_3}d_{hBN}}$, which is the effective permittivity of the compound dielectric\cite{Tuncer2002}.

\textbf{Dynamic characterization set-up}. A Vector network analyzer (VNA) is used to drive the modulator and measure its frequency response. One port of the VNA is connected to the modulator via RF cables connected to an RF on-chip probe contacting the device. The DC bias setting the modulator working point is combined in the RF path using a bias-tee. An additional RF probe connected to a 50$\Omega$ resistance is used to match the 50$\Omega$ input impedance and minimize back-reflections. A photodiode (Finisar XPD 3120R) with known frequency response (BW$\sim$70GHz) is used to collect the modulated signal, connected to the second port of the VNA. We perform a VNA calibration to account for cable losses.

The setup used for the digital data transmission measurements is similar to that used to measure BW, Fig.\ref{fig6:EO-response2}\textbf{a}. We remove the VNA and replace the first port of the VNA with a 100Gs DAC (Micram DAC4), delivering NRZ signals up to 80Gbit/s. The peak-to-peak output voltage of the DAC is 0.5V. To increase this, a RF amplifier with$\sim$50GHz BW and 1dB compression point$\sim$20dBm is used (Centellax UA1L65VM). A 6dB attenuator is then inserted between the bias tee and the GS probe. This is equivalent to connecting a 50$\Omega$ load to ground, and it is necessary to obtain impedance matching between the amplifier output port and the modulator. A pseudo-random-binary sequence (PRBS), which is a bit pattern of size $n$ that repeats after $2^n-1$ bits \cite{Proakis2008}, is prepared and encoded using a NRZ modulation format. The sequence is then generated by the DAC and applied to the modulator. The resulting optical modulation is revealed by the photodiode, and acquired using a sampling oscilloscope with electrical bandwidth$\sim$65GHz. 

\section{Acknowledgments}
We acknowledge funding from EU Graphene Flagship, ERC Grants Hetero2D, GIPT, EU Grants GRAPH-X, CHARM, EPSRC Grants EP/K01711X/1, EP/K017144/1, EP/N010345/1, EP/L016087/1, EP/V000055/1, EP/X015742/1, EP/T019697/1.

\end{document}